\documentclass[titlepage,12pt]{utarticle}
\usepackage{bbm}
\usepackage{amsmath}
\usepackage{amsfonts}
\numberwithin{equation}{section}

\begin{document}

\preprint{
  CERN--PH--TH/2006-073\\
  {\tt hep-th/0604189}\\
}

\title{\vskip-1cm Matrix Factorizations, Minimal Models \\  and Massey Products}
\author{Johanna Knapp \and Harun Omer}
\oneaddress{Department of Physics, CERN\\
     Theory Division,
     CH-1211 Geneva 23,
     Switzerland\\
    {\tt firstname.lastname@cern.ch}\\{~}\\}
 \nobreak
\Abstract{We present a method to compute the full non--linear deformations of matrix factorizations for ADE minimal models.
This method is based on the calculation of higher products in the
cohomology, called Massey products. The algorithm yields a polynomial
ring whose vanishing relations encode the obstructions of the deformations
of the D--branes characterized by these matrix factorizations. This
coincides with the critical locus of the effective superpotential which
can be computed by integrating these relations. Our results for
the effective superpotential are in agreement with those obtained
from solving the A--infinity relations. We point out a relation to the superpotentials of Kazama--Suzuki models. We will illustrate our findings by various examples, putting emphasis on the $E_6$ minimal model.
}

\date{April 2006}
\maketitle
\tableofcontents
\pagebreak

\section{Introduction and Summary}
It is by now well--known that topological D--branes for the B--model are characterized by matrix factorizations of the Landau--Ginzburg superpotential \cite{Kapustin:2002bi,Brunner:2003dc,Kapustin:2003ga,Kapustin:2003rc}. In the mathematics literature the problem is approached from the category theoretic point of view \cite{eisenbud,kontsevich,Orlov:2003yp,yoshino}. The purpose of this paper is to discuss methods to compute the effective superpotential for $B$--type Landau--Ginzburg models with boundary.\\
The effective superpotential $\mathcal{W}_{eff}$ can be understood in various ways. In $\mathcal{N}=1$ string compactifications with D--branes it can be interpreted as the four dimensional space--time superpotential. This description is valid for critical string theories.
Furthermore, the effective superpotential $\mathcal{W}_{eff}$ represents the generating functional for open string disk amplitudes. Once all the amplitudes are known, they can be integrated to give the effective superpotential. The values of the amplitudes are constrained by worldsheet consistency constraints they have to satisfy. In the case without D--branes the constraints are the WDVV equations \cite{Dijkgraaf:1990dj}. Generalizing to worldsheets with boundary, these constraints have to be extended. This was done in \cite{Lazaroiu:2000rk} for the case without integrated insertions while the general case with insertions of integrated operators was derived in \cite{Herbst:2004jp}, where it was found that amplitudes have to satisfy the $A_{\infty}$--relations. Correlators with bulk insertions have to satisfy the bulk--boundary crossing constraint in addition. The CFT Cardy constraint of \cite{Herbst:2004jp}, however, does
not necessarily hold due to possible anomalies. We will demonstrate that it
indeed breaks down for all but the simplest problems. 
A further sewing constraint for the amplitudes are the quantum
$A_{\infty}$--relations~\cite{Herbst:2006kt,Herbst:2006nn} but we will not use them in this
work. The focus of this paper is the derivation of $\mathcal{W}_{eff}$ by means of
deformation theory.\\
An interesting aspect of the effective superpotential is the fact that it encodes the obstructions of the deformations of D--branes. Turning on generic deformations usually leads away from the critical point and the problem cannot be approached within a CFT context. It is thus not surprising that the deformation problem has not yet been considered in a systematic way in the physics literature. First steps towards understanding this problem were made in \cite{Hori:2004zd,Hori:2004ja,Herbst:2004zm}. In \cite{Ashok:2004xq} some interesting results were found for the quintic. See also \cite{Brunner:2006tc} for a recent account of the deformation problem in the context of $K3$ surfaces. In the field of mathematics, however, deformation theory of matrix factorization is an active area of research. It is the aim of this work to present methods known to mathematicians in the context of deformation theory, extend them and use them to calculate $\mathcal{W}_{eff}$ for a number of examples. It seems ironic that mathematics provides efficient and elegant tools to calculate deformations of matrix factorizations, whereas the existence of an effective superpotential seems to be unknown in the mathematics literature in that context.\\
In our discussion we will mostly deal with the ADE minimal models. In these models all the possible deformations are massive and there are no marginal deformations. Consequently, it is possible to work in a purely algebraic setup which simplifies matters significantly. So far, various aspects of the A-- and D-- series minimal models have been discussed in the literature \cite{Brunner:2003dc,Kapustin:2003rc,Herbst:2004jp,Brunner:2005pq}. The classification of the matrix factorizations for the minimals model was solved some time ago \cite{greuel-knoerrer}. The complete spectra for all the ADE minimal models were given in \cite{Kajiura:2005yu}, where also the relation to the Dynkin diagrams of the simply laced Lie algebras was discussed. See also \cite{yoshino}. The mathematical framework for the deformation theory of matrix factorizations was established in \cite{laudal1, laudal2}. An explicit method to calculate deformations of matrix factorizations was presented in \cite{siqvelandPHD, Siqveland1}. In the mathematics literature the deformation problem is sometimes referred to as ``the method of computing formal moduli of modules''. The idea of the method is to find the most general deformation of a matrix factorization such that the factorization condition still holds. This can be done by an iterative procedure which amounts to calculating higher cohomology elements of the complex defined by the BRST operator $Q=\left(\begin{array}{cc}0&E\\J&0\end{array}\right)$, where $W=E\cdot J=J\cdot E$ is the matrix factorization. The central object in these calculations is the Massey product which is the generalization of the cup product to higher cohomologies. We will give a detailed description of the algorithm in section \ref{masseysec}. The spectrum of $Q$ is graded, $H(Q)=H^e\oplus H^o$, where the superscripts denote even and odd states. The odd states give the deformations, the even states the obstructions. Associating to every odd state in $H^o$ a deformation parameter $u$ the deformed $Q$--operator has the following form:
\begin{equation}
Q_{def}=Q+\sum_{\vec{m}}\alpha_{\vec{m}}u^{\vec{m}},
\end{equation}
where $u^{\vec{m}}=u_1^{m_1}\ldots u_r^{m_r}$ with $r=\mathrm{dim}\;H^o$ and $\alpha_{\vec{m}}$ are matrices which can be calculated with the Massey product algorithm. Whenever a Massey product is non--zero it contributes to $Q_{def}$ through $\alpha_{\vec{m}}$ unless it lies in $H^e$. In that latter case it contributes to the obstructions which are represented by a polynomial ring: $k[[u_1,\ldots,u_r]]/(f_1,\ldots,f_r)$, where the vanishing relations $f_i$ are polynomials in the deformation parameters $u_i$, which give precisely the critical locus of $\mathcal{W}_{eff}$: $f_i\sim\frac{\partial\mathcal{W}_{eff}}{\partial u_i}$. That these relations can be integrated to a superpotential seems a priori not obvious, but turns out to be the case for all examples we investigated. The deformed $Q$--operator then squares to the Landau--Ginzburg superpotential up to these vanishing relations. Further interesting mathematics references on this subject are \cite{Siqveland2,Siqveland3,eriksen,eriksen2,eriksen3}.\\
Without bulk insertions the effective superpotential agrees with the result obtained from the $A_{\infty}$--relations. We will show that it is possible to incorporate bulk deformations into the algorithm. In this case, however, the results disagree with those obtained from the consistency constraints. This may be an indication that there are terms missing in the bulk--boundary crossing constraint \cite{getzlerpriv}.\\
In addition we will point out a close connection between {\em effective} superpotentials associated to rank two matrix factorizations and the {\em LG} superpotentials of  Kazama--Suzuki type coset models \cite{Lerche:1991re,Eguchi:1996nh,Lerche:1996an,Eguchi:2001fm}.
For all tested cases our results agree with these superpotentials up to field redefinitions and we will present a general reason why this is so.\\

The paper is organized as follows: The next section is devoted to explaining the algorithm to compute deformations of matrix factorizations and the effective superpotential. Some explicit examples will be given. Using the example of the $E_6$ minimal model, we will demonstrate in section \ref{bulksec} how the Massey product formalism can be extended to incorporate bulk deformations. Section \ref{slohssec} is devoted to the relation between the boundary superpotential and the superpotentials obtained from coset models. In section \ref{cftsec} we give a summary of the techniques necessary to calculate the the effective superpotential using consistency constraints for the open string amplitudes. We will then merge both of the discussed methods in order to calculate the effective superpotential for the $E_6$--example and compare the results. In section \ref{moresec} we give further examples for superpotentials for minimal models of type $E$. Section \ref{conclusionsec} is devoted to conclusions and open questions. Finally, in the appendix we summarize the data for the $E_6$ model and give some new results for the $A$ minimal models, which is the only case where the Cardy constraint works.

%%%%%%%%%%%%%%%%%%%%%%%%%%%%%%%%%%%%%%%%%%%%%%%%%%%%%%%%%%%%%%%%%%%%%%%%%%%%%%
\subsection{Matrix Factorizations}
For a given $ADE$ superpotential $W$ one must first find its matrix
factorizations $W\:{\bf 1}_{N\times N}=E \cdot J=J \cdot E$.
Two matrix factorizations $(E,J)$
and $(E',J')$ are called equivalent if they are related by a similarity
transformation:
\begin{equation}
E'=U_1EU_2^{-1}\qquad J'=U_2JU_1^{-1},
\end{equation}
where $U_1,U_2\in GL(N,\mathbb{R})$ are invertible matrices with polynomial
entries. We consider only reduced factorizations,
i.e. $E(0)=J(0)=0$.
The matrix factorizations for the ADE singularities were already completely
classified by mathematicians in \cite{greuel-knoerrer}. Now,
having a set of factorizations which we label by capital letters, one constructs
a BRST operator,
\begin{equation}
Q^A=\left(
\begin{array}{cc}
0&E^A\\
J^A&0
\end{array}
\right),
\end{equation}
These operators define a graded differential $d$ by
\begin{equation}
d\psi=Q^A\psi^{AB}-(-1)^{|\psi|}\psi^{AB}Q^B.
\end{equation}
The physical states lie in the cohomology $H(d)=\frac{\mathrm{Ker}(d)}{Im(d)}$ of the
differential $d$. In the category theoretic description,
$d$ is the differential of a differential graded category and the fermionic
states correspond to $\mathrm{Hom}^1(A,B)\cong\mathrm{Ext}^1(A,B)$, the bosonic states are in $\mathrm{Hom}^2(A,B)\cong\mathrm{Ext}^2(A,B)$.\\
A method for computing the $R$--charges $q_{\psi^{AB}}$ of an open string
$\psi^{AB}\in H(d)$ stretching between the branes $A$ and $B$ was presented
in~\cite{Walcher:2004tx}; they are determined by the equation
\begin{equation}
E\psi^{AB}+R^A\psi^{AB}-\psi^{AB}R^B=q_{\psi^{AB}}\psi^{AB},
\end{equation}
where
\begin{equation}
E=\sum_i q_ix_i\frac{\partial}{\partial x_i}\quad\textrm{and}\quad W(e^{i\lambda q_i}x_i)=e^{2i\lambda}W(x_i)\:\:\forall\lambda\in\mathbb{R}.
\end{equation} 
The defining equation for the matrix $R^A$ is
\begin{equation}
EQ^A+[R^{A},Q^A]=Q^A.
\end{equation}

%%%%%%%%%%%%%%%%%%%%%%%%%%%%%%%%%%%%%%%%%%%%%%%%%%%%%%%%%%%%%%%%%%%%%%%%%%
\section{Massey Products}
\label{masseysec}
We will demonstrate that an effective superpotential can be
derived by calculating Massey products. That is the
way a mathematician would approach this problem and we will
introduce it in this paper in a form digestible to physicists.\\
In particular, an easy example will be calculated, showing that computing
deformations by this method is not too hard in practice although the 
mathematical framework needed for a rigorous treatment is quite extensive. The
method has the advantage, that a general, straightforward algorithm exists, at
least for the effective superpotentials of simple singularities.
The calculation not only gives the superpotential, but as a by-product also the
full deformed matrices of the factorization. This allows to double-check that
the result is indeed a valid deformation, minimizing the risk of an error during
the computation. The entire algorithm is already implemented in the
{\sc Singular}--package~\cite{GPS05,GPdeform} to which the tedious work can
be outsourced. The whole formalism is, however, restricted to compute only
fermionic deformations.\\
The inclusion of bulk deformations into the algorithm has not yet been done
by mathematicians and we will extend the algorithm in this paper in order to include 
them.
%%%%%%%%%%%%%%%%%%%%%%%%%%%%%%%%%%%%%%%%%%%%%%%%%%%%%%%%%%%%%%%%%%%%%%%%%%%%%%%%%%%%%%
\subsection{Mathematics}
We will review deformations only very briefly and start in the context of
differential graded Lie algebras (DGLAs). For a rigorous
treatment of this and the slightly involved definitions of Massey products
in this context we refer to the work of mathematicians~\cite{fuchs,fialowski,Penkava:1994mu}.\\\\
 A DGLA is a $\mathbb{Z}_2$- or $\mathbb{Z}$-graded  vector space $V$ over
a field $\mathbb{K}$ with a commutator of degree zero and a differential $d:V\rightarrow V$
of degree $1$, satisfying the conditions
\begin{eqnarray}
\begin{array}{l}
[\alpha,\beta]=-(-1)^{|\alpha| |\beta|}[\beta,\alpha],\\
d[\alpha,\beta]=[d\alpha,\beta]+(-1)^{|\alpha|}[\alpha,d\beta],\\
\lbrack \lbrack \alpha,\beta \rbrack,\gamma\rbrack+ (-1)^{|\alpha|(|\beta|+|\gamma|)}[[\beta,\gamma],\alpha]+
(-1)^{|\gamma|(|\alpha|+|\beta|)}[\gamma,\alpha,\beta]=0.
\end{array} 
\end{eqnarray}
A {\it formal deformation} of $V$ is defined as the power series
\begin{equation}
[g,h]_u=[g,h]+\sum_{i=1}^{\infty}u^i(\alpha_i(g,h)+\theta \beta_i(g,h)),
\end{equation}
where $\theta$ is an odd parameter. Requiring the bracket to remain bilinear
and antisymmetric and to fulfill the Jacobi identity places constraints on the
$\alpha_i$ and $\beta_i$. Deformations exist, if certain conditions of Massey products
are fulfilled and the deformations can be calculated explicitly from the Massey products.\\\\
It is not necessary to restrict to just a single parameter $u$, i.e. $L_u$ as a
deformation of an algebra over $k[u]$; we can also generalize this to $k[u_1,u_2,...,u_r]$.
Such deformations are called versal deformations and they are defined as follows.\\
{\it Definition.} A deformation $L_{R}$ of a Lie algebra $L$ parametrized by a local finite dimensional algebra $R\in\mathcal{C}$, where $\mathcal{C}$ is the category of complete local algebras,  
is a versal deformation, if for any $L_A$ parametrized by $A\in\mathcal{C}$ there is
a morphism $f:R\rightarrow A$ such that\\
(i) $L_R\otimes_R A \cong A$,\\
(ii) the map $m_R/m_R^2 \rightarrow m_A/m_A^2$, where the $m$ are the respective maximal ideals,  induced by $f$ is unique.\\\\

The whole procedure can be extended to $A_\infty$ algebras by a simple redefinition
of the commutator (see~\cite{Penkava:1994mu}):
\begin{equation}
\{\phi,\psi\}=(-1)^{(k-1)|\psi|}[\phi,\psi].% \qquad\phi\in C^k(V),\;\;\psi\in C^l(V).
\end{equation}
The $A_\infty$ structure is defined by an odd element $Q$ satisfying $\{Q,Q\}=0$ and
the action of the differential $d$ by $d\phi=\{Q,\phi\}$. The formal deformation of
such an $A_\infty$ algebra $V$ is still given by
\begin{equation}
Q_u=Q+\sum_{i=1}^{\infty} u^i(\alpha_i+\theta \beta_i), 
\end{equation}
with the condition $\{Q_u,Q_u\}=0$.\\\\
Dealing with matrix factorizations involves modules, which can be treated
in a similar manner. There, the object of interest is the infinitesimal deformation
functor of a module $M$ over a $k$-algebra $A$, defined as
\begin{equation}
\mbox{Def}_M(S)=\{(M,\theta)|M \mbox{ an } A\otimes_R S-\mbox{module, flat over }S,
M\otimes_S R \cong^{\theta} M\}/\cong.\nonumber 
\end{equation}
For finite-dimensional Ext$^1(M,M)$, Schlessingers theorem~\cite{Schlessinger} ensures the existence
of a hull $\hat H_M$ of the deformation functor, which has also been
called the {\it formal moduli} of $M$, as well as the existence of a formal versal family.
This has been studied extensively in~\cite{siqvelandPHD,Siqveland1} and we refer
to these references for proofs, details and the explicit algorithm. We restrict
ourselves to give a less rigorous down-to-earth reasoning in the next section
to make the algorithm plausible.
Afterwards an explicit example will be calculated for illustration.
%%%%%%%%%%%%%%%%%%%%%%%%%%%%%%%%%%%%%%%%%%%%%%%%%%%%%%%%%%%%%%%%%%%%%%%%%%%%%%%%%%%%%
\subsection{The Idea}
Suppose we have a matrix factorization $Q^2=W\mathbbm{1}$ of a
superpotential $W$. A deformation $Q_{def}$ of the original module $Q$ can always
be written in the form
\begin{equation}
Q_{def}=Q+\sum_{\vec{m}}\alpha_{\vec{m}} u^{\vec{m}}.\label{eq:Qdef}
\end{equation}
Here, we consider only fermionic deformations and leave the bosonic ones for
future work.
Just like $Q$, the $\alpha_{\vec{m}}$ are modules and the deformation is
parametrized by $u_1,...,u_r$ where we used the notation
\begin{equation}
u^{\vec{m}}=u_1^{m_1} u_2^{m_2}...u_d^{m_d},\;\;\vec{m}\in \mathbb{N}^d
\end{equation}
for convenience. Squaring $Q_{def}$, and comparing with $W$, we find
\begin{equation}
\sum_{\vec{m}}(Q \alpha_{\vec{m}}+\alpha_{\vec{m}} Q) u^{\vec{m}}
+\sum_{{\vec{m}_1,\vec{m}_2}}\alpha_{\vec{m}_1}\alpha_{\vec{m}_2}u^{\vec{m}_1+\vec{m}_2}=0, \label{eq:iterationeq}
\end{equation}
which must be valid at all orders of $u$. This constraint can be solved iteratively.
At order $|\vec{m}|=\sum_{i=1}^N m_i=1$ the condition $Q \alpha_{\vec{m}}+\alpha_{\vec{m}} Q=0$ holds, which means
that the basis for the odd $\alpha_{\vec{m}}$ with $|\vec{m}|=1$ is precisely the basis
for the odd cohomology.\\
The $\alpha_{\vec{m}}$ for some order $n-1$ now determine those of order $|\vec{m}|=n$ as
follows. We define
\begin{equation}
y(\vec{m}):=\sum_{\vec{m}_1+\vec{m}_2=\vec{m}}\alpha_{\vec{m}_1}\alpha_{\vec{m}_2},
\label{eq:massey}
\end{equation}
neglecting a subtlety for a moment. The $y(\vec{m})$ are called Massey Products.
For later convenience we define $\beta_{\vec{m}}=y(\vec{m})$.\\\\
All we would need in order to satisfy Eq.~(\ref{eq:iterationeq}) at order $|\vec{m}|=n$ is an arbitrary
$\alpha_{\vec{m}}$ such that
\begin{equation}
d\alpha_{\vec{m}}=Q \alpha_{\vec{m}}+\alpha_{\vec{m}} Q=-\beta_{\vec{m}}. \label{eq:iteration}
\end{equation}
Since we started at lowest order with $\alpha_{e_i}$ spanning  a basis of
Ext$^1(M,M)$, all $\beta_{\vec{m}}$ will lie either in Ext$^2(M,M)$
or be a polynomial multiple of an element lying in Ext$^2(M,M)$. In the latter case,
a matrix $\alpha_{\vec{m}}$ satisfying Eq.~(\ref{eq:iteration}) can always
be found and this matrix will again be an element of Ext$^1(M,M)$.
In the former case, where $\beta_{\vec{m}}$ lies in the even cohomology, 
a counterterm cancelling the Massey product does not exist, therefore
Eq.~(\ref{eq:iteration}) holds only mod Ext$^2(M,M)$.\\
In each iteration step, such non-cancelling terms add up so that for a
matrix factorization with dim$_k$ Ext$^2(M,M)=r$, there are $r$ polynomials
$f_i\in k[u_1,...,u_d]$, each associated with a basis element $\phi_i \in$ Ext$^2(M,M)$,
so that $Q_{def}^2=W+\sum_i f_i \phi_i$.\\
For the deformed factorization to be exactly equal to $W$, the $f_i$ must
vanish and the hull $\hat{H}_M$ of the deformation is therefore,
\begin{equation}
\hat{H}_M\simeq k[[u_1,\ldots,u_d]]/(f_1,\ldots f_r).
\end{equation}
The building up of the $f_i$ must be kept track of during the iteration.
At each order $n$, we add the new terms to the $f_i$,
\begin{equation}
f_i^1=0,\qquad f_i^n=\sum y_i^*(\langle x^*;\vec{m}\rangle)u^{\vec{m}}+f_i^{n-1},
\end{equation}
where the expression $y_i^*(\langle x^*;\vec{m}\rangle)$ denotes the proportionality
constant between $y(\vec{m})$ and the appropriate $\phi_i$.\\\\
Up to now, we have neglected one important subtlety. Namely, the ring relations
$k[u_1,...,u_d]/f_i^n$ must also be applied to the $y(\vec{m})$ and the $\beta_{\vec{m}}$,
so the above equations have to be modified. As an ingredient to be able to do
this we need to define appropriate bases $B_i$ at each order. We start with
\begin{equation}
\bar B_1=\{\vec{n} \in \mathbb{N}^d | |n|\le 1\}, \qquad
B_1=\{\vec{n} \in \mathbb{N}^d | |n|=1\}.
\end{equation}
at lowest order; in general, the $B_i$ denote the bases\footnote{In the following two equations $\vec{m}$ denotes the maximal ideal, in accordance with the notation of \cite{Siqveland1}.} for
\begin{equation}
\vec{m}^i/ \vec{m}^{i+1} + \vec{m}^i \bigcap \vec{m}(f_1^{i-1},...,f_r^{i-1}).
\end{equation}
At each order, we take
\begin{equation}
\beta_{\vec{k},\vec{l}}\in k[\vec{u}]/(\vec{m}^{n+1}+(f_1^n,...,f_r^n)),
\end{equation}
where these $\beta_{\vec{k},\vec{l}}$ are defined by the unique relation
\begin{equation}
u^{\vec{n}}=\sum_{\vec{m}\in \bigcup B_i} \beta_{\vec{n},\vec{m}}u^{\vec{m}}.\label{eq:unique}
\end{equation}
with the sum running from $i=1$ up to the the order $n$ of the iteration step.
The new defining equation for $\beta_{\vec{m}}$ is given by
\begin{equation}
\beta_{\vec{m}}=\sum_{\vec{n}\in \bigcup B_i}
\beta_{\vec{n},\vec{m}}y(\vec{n}),   \label{eq:beta}
\end{equation}
which, of course, reduces to the previous $\beta_{\vec{m}}=y(\vec{m})$ if
$\beta_{\vec{n},\vec{m}}=\delta_{\vec{n},\vec{m}}$. That is only the case when
there is no relation for the appropriate $\vec{n}$, i.e. $u^{\vec{n}}$ or a 
term proportional to it does not appear in one of the $f_i$.
A similar correction must be made for Eq.~(\ref{eq:massey}):
\begin{equation}
y(\vec{n}):=\sum_{\vec{m}_1+\vec{m}_2=\vec{m}}
\gamma_{\vec{m},\vec{n}}\alpha_{\vec{m}_1}\alpha_{\vec{m}_2}, 
\end{equation}
where $\gamma$ is defined by
\begin{equation}
\begin{array}{l}
u^{\vec{k}}=\sum_{\vec{l}\in \bigcup B_i} \gamma_{\vec{k},\vec{l}} u^{\vec{l}}
+\sum_j \gamma_{\vec{k},j} f_j^n.
\end{array}
\end{equation}

We will now clarify the algorithm by calculating an explicit example.
\subsection{A Simple Example}
Here, the $A$-series superpotential $W=-\frac{1}{5}x^5$ shall be
calculated step by step.
The factorization $Q$ and the four basis elements of the cohomology are
given by
\begin{equation}
\begin{array}{l}
Q=\begin{pmatrix}0 & x^2\\ -1/5 x^3 & 0 \end{pmatrix},\\
\alpha_{(1,0)}=\psi_1=\begin{pmatrix}0 & -x\\ -1/5 x^2 & 0 \end{pmatrix},\\
\alpha_{(0,1)}=\psi_2=\begin{pmatrix}0 & -1\\ -1/5 x & 0 \end{pmatrix},\\
\phi_1=\begin{pmatrix}x & 0\\ 0 & x \end{pmatrix},\\
\phi_2=\begin{pmatrix}1 & 0\\ 0 & 1 \end{pmatrix}.
\end{array} 
\end{equation}
The second order Massey products are
\begin{equation}
\begin{array}{rcl}
y(2,0)&=&\alpha_{(1,0)}\alpha_{(1,0)}=\frac{1}{5} x^2 \phi_1,\\
y(0,2)&=&\alpha_{(0,1)}\alpha_{(0,1)}=\frac{1}{5} \phi_1,\\
y(1,1)&=&\alpha_{(1,0)}\alpha_{(0,1)}+\alpha_{(0,1)}\alpha_{(1,0)}=\frac{2}{5}x^2 \phi_1.
\end{array}
\end{equation}
$y(0,2)$ is the only Massey product that lies in the cohomology,
therefore $f_1^2=\frac{1}{5}\vec{u}^{(0,2)}=\frac{1}{5} u_2^2$
while $f_2^2$ remains zero.
The new basis for $\vec{m}^2/(\vec{m}^3+(u_1^2,0))$ is
\begin{equation}
B_2=\{\vec{m}\in \mathbb{N}^2: |\vec{m}|=2\}-\{(0,2)\}. 
\end{equation}
For the two elements not in the cohomology, we take
\begin{equation}
\alpha_{(2,0)}=\begin{pmatrix}0&1\\0&0\end{pmatrix}
\mbox{  and  }
\alpha_{(1,1)}=\begin{pmatrix}0&0\\-2/5&0\end{pmatrix}, 
\end{equation}
to satisfy $d\alpha_{(2,0)}=-y(2,0)$ and $d\alpha_{(1,1)}=-y(1,1)$.
A different choice is possible and corresponds to a field redefinition of the effective
potential.\\
As a basis for the third order, we choose
\begin{equation}
B_3=\{|\vec{m}\in \mathbb{N}^2: |\vec{m}|=3\}-\{(0,2)+\{(1,0),(0,1)\}\}=\{(3,0),(2,1)\}.\end{equation}
%and we can always find a basis of the same structure for higher orders:
%\begin{equation}
%B_{n}=\{\vec{m}\in \mathbb{N}^{2}: |\vec{m}|={n}\}-\{(0,2)+\vec{m}\}.
%\end{equation}
We get
\begin{equation}
\begin{array}{rcl}
y(3,0)&=&\alpha_{(2,0)}\alpha_{(1,0)}+\alpha_{(1,0)}\alpha_{(2,0)}=-\frac{1}{5}x^2 \phi_2,\\
y(2,1)&=&\alpha_{(2,0)}\alpha_{(0,1)}+\alpha_{(0,1)}\alpha_{(2,0)}
+\alpha_{(1,1)}\alpha_{(1,0)}+\alpha_{(1,0)}\alpha_{(1,1)}
=\frac{1}{5} \phi_1.
\end{array}
\end{equation}
All Massey products not listed are zero. $f_1$ at third order becomes
$f_1^3=\frac{1}{5} u_2^2+\frac{1}{5}u_1^2 u_2$ while $f_2^3$ is still zero.
The $\beta_{\vec{m}}$ are
\begin{equation}
\begin{array}{l}
\beta_{(3,0)}=y(3,0),\\
\beta_{(2,1)}=y(2,1)-y(0,2)=0,
\end{array} 
\end{equation}
according to their definition in Eq.~(\ref{eq:beta}). The relation for the
$\beta_{\vec{n},\vec{m}}$ derives here from $f_1^3=0$, according to which $\frac{1}{5}u^{(2,1)}
=-\frac{1}{5}u^{(0,2)}$, fixing therewith $\beta_{(0,2),(2,1)}=-1$ from
Eq.~(\ref{eq:unique}). The choice
\begin{equation}
\alpha_{(3,0)}=\begin{pmatrix}0&0\\1/5&0\end{pmatrix}, 
\end{equation}
satisfies $d\alpha_{(3,0)}=-\beta_{(3,0)}$.
The fourth order result is
\begin{equation}
B_4=\{|\vec{m}\in \mathbb{N}^2: |\vec{m}|=4\}-\{(0,2)+\vec{k}\}\qquad \vec{k}\in \mathbb{N}^2:|\vec{k}|=2
\end{equation}
\begin{equation}
\begin{array}{rcl}
y(4,0)&=&\alpha_{(3,0)}\alpha_{(1,0)}+\alpha_{(1,0)}\alpha_{(3,0)}
+\alpha_{(2,0)}\alpha_{(2,0)}=-\frac{1}{5} \phi_1\\
y(3,1)&=&\alpha_{(3,0)}\alpha_{(0,1)}+\alpha_{(0,1)}\alpha_{(3,0)}
+\alpha_{(2,0)}\alpha_{(1,1)}+\alpha_{(1,1)}\alpha_{(2,0)}\\
&&-(\alpha_{(1,1)}\alpha_{(0,1)}+\alpha_{(0,1)}\alpha_{(1,1)})=-\phi_2
\end{array}
\end{equation}
and
\begin{equation}
\begin{array}{l}
f_1^4=\frac{1}{5}(u_2^2+u_1^2 u_2-u_1^4),\\
f_2^4=-u_1^3 u_2.
\end{array}
\end{equation}
We now have obtained a non--vanishing contribution to $f_2$ and the basis $B_5$ now looks as follows:
\begin{equation}
B_5=\{|\vec{m}\in \mathbb{N}^2: |\vec{m}|=5\}-\left(\{(0,2)+\vec{k}\}\cup \{(3,1)+\{(1,0),(0,1)\}\}\right),
\end{equation}
where $\vec{k}\in \mathbb{N}^2:|\vec{k}|=3$.\\
At fifth order - the last non-vanishing order - we find
\begin{equation}
\begin{array}{rcl}
y(5,0)&=&\alpha_{(3,0)}\alpha_{(2,0)}+\alpha_{(2,0)}\alpha_{(3,0)}
+\alpha_{(1,1)}\alpha_{(0,1)}+\alpha_{(0,1)}\alpha_{(1,1)}=\frac{3}{5}\phi_2,\\
\\
f_1:=f_1^5&=&\frac{1}{5}(u_2^2 +u_1^2 u_2 -u_1^4),\\
f_2:=f_2^5&=&\frac{3}{5}u_1^5-u_1^3 u_2.
\end{array}
\end{equation}
Now we have all necessary data and can assemble Eq~(\ref{eq:Qdef}),
\begin{equation}
\begin{array}{l}
E_{def}=x^2-u_1 x-u_2+u_1^2,\\
J_{def}=\frac{1}{5}(-x^3-u_1 x^2-u_2 x-2 u_1 u_2 +u_1^3).
\end{array}
\end{equation}
Squaring $Q_{def}$ gives
\begin{equation}
Q_{def}^2=\mathbbm{1}\frac{1}{5}(-x^5+(u_2^2+u_1^2 u_2 - u_1^4)x+(2u_1 u_2^2-3u_1^3 u_2+u_1^5)).
\end{equation}
Using the ring relations
\begin{equation}
\begin{array}{l}
f_1=\frac{1}{5}(-u_1^4+u_1^2 u_2 + u_2^2)=0,\\
f_2=-u_1^3 u_2 +\frac{3}{5} u_1^5=0,
\end{array}
\end{equation}
we see that $Q_{def}^2$ reduces to the undeformed $W=-\frac{1}{5}x^5$, confirming that
the result is indeed a valid deformation.\\
The conditions $f_i=0$ are also called critical locus. In terms of the effective
superpotential $\mathcal{W}_{eff}$, the critical locus is given by
\begin{equation}
\mathcal{Z}_{crit}=\{u\in \mathbb{C}^d|\partial_{u}\mathcal{W}_{eff}(u)=0\}. 
\end{equation}
With no bulk insertions it is of course trivial and identical to the origin.
We will now derive the effective superpotential by using the commutativity
of the partial derivatives and charge reasoning. Using the $R$-charges (denoted
by brackets),
\begin{equation}
\begin{array}{l}
[u_i]=i/5 \qquad [\mathcal{W}_{eff}]=6/5\qquad [f_1]=4/5 \qquad [f_2]=5/5
\end{array} 
\end{equation}
we find that $[\partial_{u_1}\partial{u_2}\mathcal{W}_{eff}(u,t)]=3/5$ and therefore
\begin{eqnarray}
\begin{array}{l}
\partial{u_1}\mathcal{W}_{eff}(u;t=0)=\mathbb{R} f_2 + \mathbb{R} u_1 f_1\;\;\mbox{  and}\\
\partial{u_2}\mathcal{W}_{eff}(u;t=0)=\mathbb{R} f_1.
\end{array}
\end{eqnarray}
The coefficients can be determined by using the commutativity of the second order partial
derivatives. The effective superpotential is now
\begin{equation}
\mathcal{W}_{eff}(u;t=0)=\frac{u_1^6}{15}-\frac{u_1^4 u_2}{5}+\frac{u_1^2 u_2^2}{10}+\frac{u_2^3}{15}.\label{eq:asol1}
\end{equation} 
The result is equivalent to that in~\cite{Herbst:2004jp} and related to it
by the field redefinition $u_2\rightarrow u_2+u_1^2$.
This can be understood by remembering that during the iteration, we chose an
arbitrary element $\alpha_{\vec{m}}$ to satisfy Eq.~(\ref{eq:iteration}). Since
the operator $d$ is nilpotent, each $\alpha_{\vec{m}}$ is only fixed up to the
addition of an exact matrix.
With a specific choice of the $\alpha_{\vec{m}}$, the cited results can be reproduced
directly without need for a field redefinition.
%%%%%%%%%%%%%%%%%%%%%%%%%%%%%%%%%%%%%%%%%%%%%%%%%%%%%%%%%%%%%%%%%%%%%%%%%%%%%%%%%%%%
\subsection{The Case $E_6$}
The $E_6$ superpotential $W=x^3+y^4-z^2$ has a matrix factorization
\begin{equation}
Q=\begin{pmatrix}
0 & 0 & -y^2-z & x\\
0 & 0 & x^2 & y^2-z\\
-y^2+z & x & 0 & 0\\
x^2 & y^2+z & 0 &0
\end{pmatrix}.
\end{equation}
A basis for the cohomology is
\begin{equation}
\begin{array}{l}
\alpha_{(1,0)}=\psi_1=
\begin{pmatrix}
0 & 0 & 0 & 1\\
0 & 0 & -x & 0\\
0 & 1 & 0 & 0\\
-x& 0 & 0 & 0
\end{pmatrix},\\
\alpha_{(0,1)}=\psi_2=
\begin{pmatrix}
0 & 0 & 0 & y\\
0 & 0 &-xy & 0\\
0 & y & 0 & 0\\
-xy & 0 & 0 & 0
\end{pmatrix},\\
\phi_1=\mathbbm{1},\\
\phi_2=y\mathbbm{1}.
\end{array}
\end{equation}
The full deformation is found to be
\begin{equation}
\hat{H}_M\simeq k[[u_1,u_4]]/(u_4^3-\frac{3}{4} u_4 u_1^8 -\frac{5}{64}u_1^{12},
\;\;3 u_4^2 u_1+\frac{3}{2} u_4 u_1^5+\frac{1}{8}u_1^9),
\end{equation}
and explicitly as
\begin{eqnarray}
E_{def}=
\begin{pmatrix}
-y^2+z+\frac{u_1^3}{2} y+\frac{u_1^6}{8}+\frac{3}{2}u_4 u_1^2
& x+u_1 y + u_4\\
x^2-u_1xy +u_1^2 y^2 + 2 u_4 u_1 y-u_4 x + u_4^2
& y^2+z -\frac{u_1^3}{2} y-\frac{u_1^6}{8}-\frac{3}{2}u_4 u_1^2
\end{pmatrix},\nonumber\\
J_{def}=\begin{pmatrix}
-y^2-z+\frac{u_1^3}{2} y+\frac{u_1^6}{8}+\frac{3}{2}u_4 u_1^2
&x+u_1 y+u_4\\
x^2- u_1 xy+u_1^2y^2+2 u_4 u_1 y-u_4 x+u_4^2
&y^2-z-\frac{u_1^3}{2} y-\frac{u_1^6}{8}-\frac{3}{2}u_4 u_1^2
\end{pmatrix}.
\end{eqnarray}
Note that we have slightly changed notation and labelled the deformation parameters by their $R$--charge.
Squaring $Q_{def}$ gives
\begin{eqnarray}
J_{def}E_{def}=E_{def}J_{def}=W(x,y,z)+f_2 y+ (f_1+\frac{1}{2}u_1^3f_2),
\end{eqnarray}
proving that it is indeed a deformation.
The corresponding effective superpotential is
\begin{equation}
\mathcal{W}_{eff}(u)=u_4^3u_1+\frac{3}{4}u_4^2u_1^5+\frac{1}{8}u_4 u_1^9+\frac{5}{832}u_1^{13}.
\end{equation}
%%%%%%%%%%%%%%%%%%%%%%%%%%%%%%%%%%%%%%%%%%%%%%%%%%%%%%%%%%%%%%%%%%%%%%%%%%%%%%%%%%%
We can check this result by using of the computer algebra package {\sc Singular} which can also perform
this calculation. The output gives the full deformed matrices as well as the
critical locus. All that remains to do is to integrate a combination of the latter
to a superpotential.
\section{Bulk Deformations}
\label{bulksec}
Instead of restricting ourselves to deformations of matrices whose product
is again the superpotential $W$, we now allow for perturbations of $W$ as well.
%%%%%%%%%%%%%%%%%%%%%%%%%%%%%%%%%%%%%%%%%%%%%%%%%%%%%%%%%%%%%%%%%%%%%%%%%%%%%%%%
\subsection{Polynomial Division}
Polynomial division was used in~\cite{Herbst:2004zm} to derive the bulk deformations
of the $A$ model.
The polynomial division gives
\begin{equation}
J_{def}(x,u,t)=\frac{W(x,t)}{E_{def}(x,u)}=\frac{1}{5}(x^3+u_1 x^2+(u_1^2+u_2-5 t_2)x+2 u_1 u_2 - 5t_3
-5 t_2 u_1 + u_1^3) + r, 
\end{equation}
where $W(x,t)$ is the deformed bulk superpotential.
Vanishing of the remainder $r$,
\begin{eqnarray}
\begin{array}{l}
r=r_1(u,t)x+ r_2(u,t),\\
r_1(u,t)= t_2 t_3-t_5 - t_3 u_2 -t_2 u_1 u_2+\frac{u_1^3 u_2}{5}+\frac{2 u_1 u_2^2}{5},\\
r_2(u,t)= t_2^2-t_4-t_3 u_1 - t_2 u_1^2+\frac{1}{5}u_1^4-t_2 u_2 +\frac{3}{5} u_1^2 u_2+\frac{1}{5}u_2^2,
\end{array}
\end{eqnarray}
gives the critical locus for this case. It defines the effective superpotential
by 
\begin{eqnarray}
\begin{array}{l}
\partial_{u_1}\mathcal{W}_{eff}(u;t)=r_1(u,t)+r_2(u,t) u_1,\\
\partial_{u_2}\mathcal{W}_{eff}(u;t)=r_2(u,t),
\end{array}
\end{eqnarray}
Of course, the $A$ model is special, in that it is the only factorization
with $1\times 1$ matrices and for higher dimensional matrix factorizations
we would have to take the determinant of the equation. The graver problem is that
polynomial division does not necessarily have a solution. The $E_6$ model with
the superpotential derived earlier is an example for that. This is not too surprising
since the polynomial division implies that only either $J$ {\it or} $E$ have
bulk deformations whereas we have chosen the boundary deformation to be
symmetrical in $E$ and $J$ in the $E_6$ computation. Therefore, polynomial
division generally requires at least
non-trivial guesswork in starting with a suitable boundary potential, 
and normally a solution does not even exist. 
%%%%%%%%%%%%%%%%%%%%%%%%%%%%%%%%%%%%%%%%%%%%%%%%%%%%%%%%%%%%%%%%%%%%%%%%%%%%%%%%%%
\subsection{Adapting the Massey Product Method}
The method of computing formal moduli can be adapted to incorporate bulk insertions.
It shall be demonstrated again at hand of the $E_6$ example. The most general
deformation for this singularity is obtained by deforming
with all elements in the chiral ring,
\begin{equation}
W_{def}(x,s)=W(x)-s_2 xy^2-s_5 xy-s_6 y^2 - s_8 x - s_9 y - s_{12}. 
\end{equation}
In the physical theory the $s_i$ are functions of bulk parameters $t_i$ as 
given in~\cite{Klemm:1991vw}. Parallel to the reasoning in computing the
formal moduli, we are looking for a deformation
\begin{equation}
\tilde Q=Q_{def}+\sum_{\vec{m}} \tilde \alpha_{\vec{m}}s^{\vec{m}},
\end{equation}
which squares to $W(x,s)$. The analogue to Eq.~(\ref{eq:iterationeq}) becomes
\begin{equation}
\sum_{\vec{m}}(Q_{def} \tilde\alpha_{\vec{m}}+\tilde\alpha_{\vec{m}} Q_{def}) s^{\vec{m}}
+\sum_{\vec{m}_1,\vec{m}_2}\tilde\alpha_{\vec{m}_1}
\tilde\alpha_{\vec{m}_2}s^{\vec{m}_1+\vec{m}_2}=W_{def}(x,s)-W(x).
\end{equation}
The $s_{12}$- and $s_8$-terms lie in Ext$^2(E,E)$ and are therefore of no concern.
They will simply be subtracted from the $f_i'$ in $Q_{def}^2=W+f_1'y+f_2'$.
For the other four deformations we chose odd $\tilde \alpha_{e_i}$ such that
\begin{equation}
\begin{array}{rcl}
\left[Q_{def},\tilde\alpha_{e_1}\right]&=&-x,\\
\left[Q_{def},\tilde\alpha_{e_2}\right]&=&-y^2,\\
\left[Q_{def},\tilde\alpha_{e_3}\right]&=&-xy,\\
\left[Q_{def},\tilde\alpha_{e_4}\right]&=&-xy^2.
\end{array}
\end{equation}
Possible choices are
\begin{eqnarray}
\begin{array}{rcl}
\tilde\alpha_{(1,0,0,0)}&=&-1
\begin{pmatrix}0&0&0&0\\0&0&1&0\\0&0&0&0\\1&0&0&0\end{pmatrix},\\
\tilde\alpha_{(0,1,0,0)}&=&\frac{1}{2}
\begin{pmatrix}0&0&1&0\\0&0&0&-1\\1&0&0&0\\0&-1&0&0\end{pmatrix},\\
\tilde\alpha_{(0,0,1,0)}&=&y\tilde\alpha_{(1,0,0,0)}-u_1\tilde\alpha_{(0,1,0,0)},\\
\tilde\alpha_{(0,0,0,1)}&=&\left(\frac{1}{8}u_1^4+\frac{1}{2}x\right)\tilde\alpha_{(0,1,0,0)}\\
&&+\begin{pmatrix}
0&0&0&\frac{1}{4}u_1^2\\
0&0&-(u_2 u_4^2+\frac{u_1^6}{8})-\frac{1}{4}u_1^2x&0\\
0&\frac{1}{4}u_1^2&0&0\\
-(u_1^2 u_4+\frac{u_1^6}{8})-\frac{1}{4}u_1^2x&0&0&0\end{pmatrix}.
\end{array}\nonumber
\end{eqnarray}
At second order we find,
\begin{eqnarray}
\begin{array}{rcll}
y(0,0,0,2)&=&(\frac{1}{4}x^2+\frac{1}{16}x)\mathbbm{1}&\mbox{ mod Ext}^2(E,E),\\
y(0,1,0,1)&=&\frac{1}{2}x\mathbbm{1}&\mbox{ mod Ext}^2(E,E),\\
y(0,0,1,1)&=&-\frac{1}{2}u_1 x\mathbbm{1}&\mbox{ mod Ext}^2(E,E),\\
\end{array}
\end{eqnarray}
all others are zero (mod Ext$^2(E,E)$). The associated second order $\tilde\alpha$'s are,
\begin{eqnarray}
\begin{array}{rcl}
\tilde\alpha(0,0,0,2)&=&(\frac{1}{16}u_1^4-\frac{1}{4}u_2+\frac{1}{4}x)\tilde\alpha(1,0,0,0)-\frac{1}{4}u_1\tilde\alpha(0,0,1,0),\\
\tilde\alpha(0,1,0,1)&=&\frac{1}{2}\tilde\alpha(1,0,0,0),\\
\tilde\alpha(0,0,1,1)&=&\frac{1}{2}u_1\tilde\alpha(1,0,0,0).\\
\end{array}
\end{eqnarray}
At third (and last order) only one term remains,
\begin{equation}
\begin{array}{rcl}
y(0,0,0,3)&=&\frac{1}{16}u_1^2x\mathbbm{1},\\
\tilde\alpha(0,0,0,3)&=&\frac{1}{16}u_1^2\tilde\alpha(1,0,0,0).
\end{array}
\end{equation}
Adding these terms up to get the complete deformed factorization $\tilde Q$
and squaring,
the result is $W_{def}(x,s)+\tilde f_1 y+\tilde f_2$ with some $\tilde f_1$ and $\tilde f_2$.
The relations
\begin{equation}
\begin{array}{rcl}
\partial_{u_1}\mathcal{W}_{eff}(u;s)&:=&\tilde f_1,\\
\partial_{u_2}\mathcal{W}_{eff}(u;s)&:=&\left(\frac{1}{2}u_1^3\tilde f_2+\tilde f_1\right),
\end{array}
\end{equation}
allow to integrate to the full superpotential,
\begin{equation}
\begin{array}{rcl}
\mathcal{W}_{Massey}(u;s)&=&\frac{5}{64\cdot 13}u_1^{13}+\frac{1}{8}u_4 u_1^9+\frac{3}{4}u_4^2u_1^5
+u_4^3 u_1 + \frac{1}{32\cdot 11}s_{2} u_1^{11}\\
&&+\frac{3}{64\cdot 9}s_{2}^2u_1^9-\frac{3}{8\cdot 8}s_5u_1^8+\frac{3}{8\cdot 7}s_6u_1^7+\frac{3}{64\cdot 7}s_{2}^3 u_1^7\\
&&+\frac{1}{16}s_{2}^2u_4 u_1^{5}
-\frac{1}{10}\left(s_8+\frac{1}{4}s_6 s_{2}\right)u_1^5-\frac{1}{2}s_5 u_4 u_1^4\\
&&+\frac{1}{2\cdot 4}s_9 u_1^4-\frac{1}{4}s_{2}u_4^2 u_1^3+\frac{1}{2}s_6 u_4 u_1^3
-\frac{1}{12}\left(s_8 s_{2}-s_5^2\right)u_1^3\\
&&+\frac{1}{4}s_5 s_{2} u_4 u_1^2-\frac{1}{4}s_6 s_5 u_1^2
+\frac{1}{4}s_{2}^2 u_4^2 u_1-\frac{1}{2}s_{2}s_6 u_4 u_1\\
&&-s_8 u_4 u_1+(s_{12}+\frac{1}{4}s_6^2)u_1-\frac{1}{2}s_5 u_4^2
+s_9u_4+\mbox{const.}
\end{array}\label{eq:fullpot}
\end{equation}
The integration constant is an arbitrary function of the $s_i$ but, of course, independent
from the $u_i$.
In the bulk theory, usually the coordinate transformation $s_i\rightarrow t_i(s_i)$
is used for the sake of adherence to the constant metric coordinate
system~\cite{Klemm:1991vw},
\begin{eqnarray}
\begin{array}{rcl}
s_{2}&=&t_{2},\\
s_5&=&t_5,\\
s_6&=&\left(t_6-\frac{t_{2}^3}{2}\right),\\
s_8&=&\left(t_8-t_6 t_{2}+\frac{t_{2}^4}{12}\right),\\
s_9&=&\left(t_9-t_5 t_{2}^2\right),\\
s_{12}&=&\left(t_0-\frac{t_6^2}{2}-\frac{t_5^2 t_{2}}{2}-\frac{t_8t_{2}^2}{2}+\frac{t_6 t_{2}^3}{6}\right).
\end{array}\label{eq:si}
\end{eqnarray}

%%%%%%%%%%%%%%%%%%%%%%%%%%%%%%%%%%%%%%%%%%%%%%%%%%%%%%%%%%%%%%%%%%%%%%%%%%%%%%%%%%%
\section{Relation to the Hermitian Symmetric Space Coset Models}
\label{slohssec}
In this section we show that the {\em effective} superpotentials associated with rank two matrix
factorizations can be related to the {\em Landau--Ginzburg} potentials of
simply--laced, level one, hermitian symmetric space (SLOHSS)
models~\cite{Lerche:1991re}\footnote{Quite recently an interesting relation between Kazama Suzuki superpotentials, matrix factorizations and knot homology was discovered \cite{khovanov1,khovanov2,Gukov:2005qp}.}. These Kazama--Suzuki type models are represented by
cosets $G/H$, where the group $G$ is divided by its maximal subgroup $H$.
In the $E_6$ case this is
\begin{equation}
\frac{E_6}{SO(10)\times U(1)}.\nonumber
\end{equation}
Landau--Ginzburg potentials for the deformed SLOHSS models were
derived in \cite{Eguchi:2001fm}. They are obtained by 
first expressing the Casimirs $V_i$ of the group $G$, in terms of the Casimirs
of $H$, which are called $x_i$. Next, we set $V(x_i)=v_i$, identifying
the Casimirs with deformation parameters $v_i$ of the superpotential.
This yields a system of equations, where the $x_i$ that appear linearly
can be eliminated. The remaining equations can then be integrated to a superpotential.
The explicit form of it was given as \cite{Eguchi:2001fm}:
\begin{eqnarray}
W(x,z,w)&=&x^{13}-\frac{25}{169}x\,z^3+\frac{5}{26}x^2\,w_2\nonumber \\
&&+z\left(x^9+x^7\,w_1+\frac{1}{3}x^5\,w_1^2-x^4\,w_2-\frac{1}{3}x^2\,w_1\,w_2+\frac{1}{12}x^3\,w_3-\frac{1}{6}x\,w_4+\frac{1}{3}w_5 \right)\nonumber \\
&&+\frac{247}{165}x^{11}\,w_1+\frac{13}{15}x^9\,w_1^2-\frac{39}{20}x^8\,w_2+\frac{169}{945}x^7\,w_1^3+\frac{13}{105}x^7\,w_3-\frac{26}{15}x^6\,w_1\,w_2\nonumber \\
&&+\frac{13}{225}x^5\,w_1\,w_3-\frac{13}{50}x^5\,w_4-\frac{91}{180}x^4\,w_1^2\,w_2+\frac{13}{30}x^4\,w_5+\frac{13}{15}x^3w_2^2-\frac{13}{90}x^3\,w_1\,w_4\nonumber \\
&& -\frac{13}{120}x^2\,w_2\,w_3+\frac{13}{90}x^2\,w_1\,w_5-\frac{13}{270}x\,w_6-\frac{13}{360}w_1^4\,w_2+\frac{13}{90}w_1^2\,w_5.
\label{eq:slohss}
\end{eqnarray}
This superpotential is, up to a (quite complicated) field redefinition and the choice
of an integration constant precisely the effective superpotential associated
to the $2\times 2$--factorization of the minimal model in our results.
The ansatz for such a field redefinition looks as follows:
\begin{equation}
u_1\rightarrow \alpha_1 x\qquad u_4\rightarrow \alpha_2z+\alpha_3x^4+\alpha_4w_1^2+\alpha_5w_1x^2
\end{equation}
\begin{eqnarray}
\begin{array}{c}
s_{12}\rightarrow\beta_1w_6+\beta_2w_4w_1^2+\beta_3w_3^2+\beta_4w_3w_1^3+\beta_5w_2^2w_1+\beta_6w_1^6 \\
s_9\rightarrow\beta_7w_5+\beta_8w_2w_1^2\qquad s_8\rightarrow\beta_9w_4+\beta_{10}w_3w_1+\beta_{11}w_1^4\\
s_6\rightarrow\beta_{12}w_3+\beta_{13}w_1^3\quad s_5\rightarrow\beta_{14}w_2\quad s_2\rightarrow\beta_{15}w_1.
\end{array}
\end{eqnarray}
Plugging this into Eq.~(\ref{eq:fullpot}) and comparing with
Eq.~(\ref{eq:slohss}) all the parameters $\alpha_i,\beta_i$ can be fixed. 
It turns our that the integration constant in Eq.~(\ref{eq:fullpot}) is a crucial
degree of freedom to get the potentials in agreement. Without it, agreement can
only be achieved by setting one deformation parameter to zero.\\\\
Note that for the $E_8$ there is no SLOHSS model. However, there is also no
rank $2$ matrix factorization of the $E_8$ Landau--Ginzburg potential.
With the formalism of matrix factorizations we can also derive superpotentials
for matrix factorizations of higher rank.\\

It thus seems that only the effective superpotentials associated to $2\times 2$ matrix factorizations have a direct connection to the superpotentials coming from SLOHSS models. It is interesting to ask why and how the matrix factorizations are encoded in these coset models. We now give a qualitative explanation of how this happens\footnote{We thank Nicholas Warner for helpful discussions.}. One way to calculate such superpotentials (at least in principle) is to eliminate as many variables as possible \cite{Eguchi:2001fm}. Take for instance the $E_6$--example:
\begin{equation}
W=x^3+y^4-z^2\;\mbox{(+ bulk deformations)}.
\end{equation}
The equation $W=0$ describes an ALE space, a two complex dimensional surface in $\mathbb{C}^3$. The idea is to start with the simplest algebraic objects and find lines and quadrics on this surface.
Using the ansatz
\begin{equation}
x=\lambda y+\alpha(\lambda),
\end{equation}
where $\lambda$ is a variable of weight one and $\alpha$ is a homogeneous polynomial of weight four,
the variable $z$ can be eliminated from $W$:
\begin{equation}
z=\sqrt{x^3+y^4\;\mbox{(+ bulk deformations)}}
\end{equation}
In this relation the ansatz for $x$ can be inserted. The equation on $z$ describes
a quadric if the expression under the square root is a perfect square. In this case
one gets
\begin{equation}
z=y^2+\gamma_1(\lambda) y+\gamma_2(\lambda).
\end{equation}
Lines and quadrics on the surface $W=0$ are then given by
\begin{eqnarray}
A_1&=&\lambda y+\alpha(\lambda)\:=\:0\nonumber \\
A_2&=&y^2+\gamma_1(\lambda) y+\gamma_2(\lambda)\:=\:0.
\end{eqnarray}
The Nullstellensatz tells us that this is consistent with $W=0$ if $W$ has the form
\begin{equation}
W=A_1B_1+A_2B_2,
\end{equation}
for some $B_1,B_2$. This is precisely a $2\times 2$ matrix factorization of the superpotential! Our results for the ADE minimal models thus suggest that there is a direct relation between the coset model LG superpotentials and $\mathcal{W}_{eff}$ for $2\times 2$ matrix factorizations of the ADE minimal models. The results are therefore consistent with the argument presented above.
Note that, at least in the presently discussed form, such a computation cannot be generalized to higher dimensional matrix factorizations.
%%%%%%%%%%%%%%%%%%%%%%%%%%%%%%%%%%%%%%%%%%%%%%%%%%%%%%%%%%%%%%%%%%%%%%%%%%%%%%%%%%%%%
\section{Using the Method of CFT Consistency Conditions}
\label{cftsec}
As we have demonstrated, the results of the different approaches are in agreement,
confirming the correctness of the method we introduced and establishing a link
between these approaches. However, there is no reason why these general solutions
should be consistent with all physical constraints. Therefore, we will now use the
CFT consistency conditions to rederive the superpotential. This approach does not
give the deformed matrices $J_{def}$ and $E_{def}$, but on the other hand one gets all the correlators.\\\\
The idea is to determine the values of the allowed correlators of the model by imposing consistency constraints. These constraints can be viewed as generalizations of the WDVV equations for the bulk \cite{Dijkgraaf:1990dj}. Adding boundaries on the worldsheet, one gets additional sewing constraints which were derived in \cite{Lazaroiu:2000rk} for the case without integrated insertions. For the case where insertions of integrated boundary states are allowed, the consistency conditions were derived in \cite{Herbst:2004jp}. In the following, we will give a short summary of the necessary steps to compute the superpotential. The input data are the matrix factorizations, the spectrum and the $R$--charges as defined in \cite{Walcher:2004tx}.

\subsection{Two-- and Three--Point Functions} 
We begin by determining the values of the boundary three--point functions and the bulk--boundary two--point functions \cite{Kapustin:2003rc,Herbst:2004ax}:
\begin{equation}
\langle\phi_i\psi_a^{AA}\rangle=\frac{1}{(2\pi i)^n}\oint \mathrm{d}^nx\frac{\phi_i\mathrm{STr}\left(\left((\partial Q^A)^{\wedge n}\right)\psi_a^{AA}\right)}{\partial_1 W\ldots\partial_nW}
\end{equation} 
Here $n$ is the number of variables in $Q$, $\phi_i$ are the elements of the bulk chiral ring $\frac{\mathbb{C}[x_i]}{\partial_i W}$, $\mathrm{STr}$ is the supertrace and the wedge product is
\begin{equation}
\left(\partial Q^A\right)^{\wedge n}=\frac{1}{n!}\sum_{\sigma\in S_n}(-1)^{|\sigma|}\partial_{\sigma(1)}Q^{A}\ldots\partial_{\sigma(n)}Q^A.
\end{equation}
The boundary three--point function is:
\begin{equation}
\langle \psi_a^{AB}\psi_b^{BC}\psi_c^{CA}\rangle=\frac{1}{(2\pi i)^n}\oint\mathrm{d}^nx\frac{\mathrm{STr}\left((\partial Q^A)^{\wedge n}\psi_a^{AB}\psi_b^{BC}\psi_c^{CA}\right)}{\partial_1 W\ldots\partial_n W}
\end{equation}

\subsection{Correlators and Selection Rules}
\label{selection-subsec}
In the second step we use a set of selection rules to find out which correlators are possibly non--vanishing. First define correlators with arbitrary numbers of bulk and boundary insertions\footnote{From now on we will suppress the indices labelling the brane.}:
\begin{eqnarray}
\label{bdef}
B_{a_0\ldots a_m;i_1\ldots i_m}&:=&(-1)^{\tilde{a}_1+\ldots\tilde{a}_{m-1}}\left\langle\psi_{a_0}\psi_{a_1}P\int\psi^{(1)}_{a_2}\ldots\psi^{(1)}_{a_{m-1}}\psi_{a_m}\int\phi^{(2)}_{i_1}\ldots \phi^{(2)}_{i_n} \right\rangle \nonumber \\
&=&-\left\langle\phi_{i_1}\psi_{a_0}P\int\psi^{(1)}_{a_1}\ldots\int\psi^{(1)}_{a_m}\int\phi_{i_2}^{(2)}\ldots\int\phi_{i_n}^{(2)}\right\rangle.
\end{eqnarray}
The integrated insertions of bulk and boundary fields are defined in \cite{Dijkgraaf:1990dj,Herbst:2004jp}. The second equality can be derived from Ward identities. Here we introduced the suspended grade of the boundary field $\psi_a$:
\begin{equation}
\tilde{a}:=|\psi_a|+1.
\end{equation} 
Furthermore we can introduce a metric on the boundary ring:
\begin{equation}
\omega_{ab}=\langle\psi_a\psi_b\rangle=(-1)^{\tilde{a}}B_{0ab}=(-1)^{\tilde{\omega}}(-1)^{\tilde{a}\tilde{b}}\omega_{ba},
\end{equation}
where the ``0'' stands for the insertion of the unit operator.
It can be used to raise and lower indices:
\begin{equation}
B^a_{\phantom{a}a_1\ldots a_m}:=\omega^{ab}B_{ba_1\ldots a_m}.
\end{equation}
The correlators (\ref{bdef}) are cyclic in the boundary insertions:
\begin{equation}
B_{a_0\ldots a_m;i_1\ldots i_m}=(-1)^{\tilde{a}_m(\tilde{a}_0+\ldots+\tilde{a}_{m-1})}B_{a_ma_0\ldots a_{m-1};i_1\ldots i_m}.
\end{equation}
Furthermore they are symmetric under permutations of the bulk indices.\\
It turns out to be convenient to define
\begin{equation}
B_{a_0a_1}=B_{a_0}=B_i=0.
\end{equation}
A correlator (\ref{bdef}) satisfies the following selection rules:
\begin{itemize}
\item Charge selection rule. \\
The $R$--charge of the correlator must be equal to the background charge. For the minimal models the background charge is given by:
\begin{equation}
q_b=\frac{k-2}{k},
\end{equation}
where $k$ is the dual Coxeter number.\\
The $R$--charges of the integrated insertions are
\begin{eqnarray}
q^I_{\psi}&=&q_{\psi}-1,\\
q^I_{\phi}&=&q_{\phi}-2.
\end{eqnarray}
For minimal models these charges are always negative.
\item The correlators must have the same suspended degree as the boundary metric:
\begin{equation}
\label{z2selection}
B_{a_0a_1\ldots a_m;i_1\ldots i_n}=0\qquad \mathrm{unless}\;\;\tilde{a}_0+\ldots+\tilde{a}_m=\tilde{\omega}.
\end{equation}
\item Insertions of the unit operator are only allowed if there are no integrated insertions.
\begin{equation}
B_{0a_1\ldots a_m;i_1\ldots i_n}=0\qquad \mathrm{for}\;\;m\geq3\;\;\mathrm{or}\;\;n\geq 1.
\end{equation}
\end{itemize}

\subsection{Consistency Conditions}
Next, we determine the correlators by imposing the generalized WDVV--constraints.
For this we introduce generating functions for the bulk perturbations, given set of boundary insertions, which satisfy the property
\begin{equation}
B_{a_0\ldots a_m;i_1\ldots i_n}=\partial_{i_1}\ldots\partial_{i_n}\mathcal{F}_{a_0\ldots a_m}(t)|_{t=0}.
\end{equation}
For $m\geq 2$ the generating functions are given by
\begin{eqnarray}
\mathcal{F}_{a_0\ldots a_m}&=&(-1)^{\tilde{a}_1+\ldots+\tilde{a}_{m-1}}\langle\psi_{a_0}\psi_{a_1}P\int\psi_{a_2}\ldots\int\psi_{a_{m-1}}\psi_{a_m}e^{\sum_pt_p\int\phi_p}\rangle\nonumber\\
&=&(-1)^{\tilde{a}_1+\ldots+\tilde{a}_{m-1}}\sum_{N_0\ldots N_{h_c-1}=0}^{\infty}\prod_{p=0}^{h_c-1}\frac{t_p^{N_p}}{N_p!}\langle\psi_{a_0}\psi_{a_1}P\int\psi_{a_2}\ldots\int\psi_{a_{m-1}}\psi_{a_m}\left[\int\phi_p \right]^{N_p}\rangle,\nonumber\\
\end{eqnarray}
where $h_c$ is the dimension of the bulk chiral ring.
For $m=0$ and $m=1$ we define $\mathcal{F}_a(t)$ and $\mathcal{F}_{ab}(t)$ by
\begin{eqnarray}
\partial_i\mathcal{F}_a(t)&=&-\langle\phi_i\psi_ae^{\sum_pt_p\int\phi_p}\rangle,\\
\partial_i\mathcal{F}_{ab}(t)&=&-\langle\phi_i\psi_aP\int\psi_be^{\sum_pt_p\int\phi_p} \rangle.
\end{eqnarray}
Now we impose the consistency constraints on the correlators. We start with the $A_{\infty}$--relations:
\begin{equation}
\label{ainfty}
\sum_{\stackrel{k,j=0}{k\leq j}}^m(-1)^{\tilde{a}_1+\ldots\tilde{a}_k}\mathcal{F}^{a_0}_{\phantom{a_0}a_1\ldots a_kca_{j+1}\ldots a_m}(t)\mathcal{F}^{c}_{\phantom{c}a_{k+1}\ldots a_j}(t)=0.
\end{equation}
The second consistency condition is the bulk--boundary crossing constraint\footnote{It is possible that there are terms missing in this relation \cite{getzlerpriv}.}:
\begin{equation}
\label{bbcr}
\partial_i\partial_j\partial_k\mathcal{F}(t)\eta^{kl}\partial_l\mathcal{F}_{a_0\ldots a_m}(t)=
\end{equation}
\[
=\sum_{0\leq m_1\leq\ldots m_4\leq m}(-1)^{\tilde{a}_{m+1}+\ldots+\tilde{a}_{m_3}}\mathcal{F}_{a_0\ldots a_{m_1}ba_{m_2+1}\ldots a_{m_3}ca_{m_4+1}\ldots a_m}\partial_i\mathcal{F}^b_{\phantom{b}a_{m_1+1}\ldots a_{m_2}}\partial_j\mathcal{F}^c_{\phantom{c}a_{m_3+1}\ldots a_{m_4}}.
\]
Here, $\mathcal{F}(t)$ is the bulk WDVV potential and $\eta^{kl}$ is the inverse of the topological metric $\eta_{kl}=\langle \phi_0\phi_k\phi_l\rangle$.\\
These two constraints alone do not determine the values of all the amplitudes. For the $A$--minimal models one can use the Cardy constraint to fix all the correlators.
The Cardy constraint was derived in~\cite{Herbst:2004jp} and takes the form
\begin{eqnarray}
\begin{array}{l}
\partial_i\mathcal{F}_{a_0...a_n} \eta^{ij}\partial_j \mathcal{F}_{b_0...b_n}=\\
\displaystyle{\sum_{\stackrel{0\le n_1 \le n_2 \le n}{0\le m_1 \le m_2 \le m}}
(-1)^{s+\tilde c_1 + \tilde c_2} \omega^{c_1 d_1} \omega^{c_2 d_2}
\mathcal{F}_{a_0...a_{n_1} d_1 b_{m_1+1}...b_{m_2} c_2 a_{n_2+1}...a_n}
\mathcal{F}_{b_0...b_{m_1} c_1 a_{n_1+1}...a_{n_2} d_2 b_{m_2+1}...b_m}}\\
\label{eq:cardy}
\end{array}
\end{eqnarray}
It turns out that this sewing constraint is only valid in the case of
the $A$--models. (We give some new results for these models in Appendix \ref{cardy-app}.)
For other minimal models the Cardy--condition turns out to be in contradiction
with (\ref{ainfty}) and (\ref{bbcr}).
This does not come as a surprise since in the derivation of the above formula it was assumed
that the annulus amplitude is metric independent, which is not necessarily true due to the
possible existence of anomalies in the $Q$--symmetry. Still, it had been hoped that
the Cardy constraint Eq.~(\ref{eq:cardy}) could nevertheless be imposed
to get the topological part of the amplitude, but our results show that this is not
the case.

\subsection{Calculation of the Effective Superpotential}
With these preparations we are now able to compute the superpotential.
When the values of all the correlators have been fixed, the effective superpotential is given by the following expression:
\begin{equation}
\mathcal{W}_{eff}(s,t)=\sum_{m\geq 1}\frac{1}{m!}s_{a_m}\ldots s_{a_1}\mathcal{A}_{a_1\ldots a_m},
\end{equation}
where
\begin{equation}
\mathcal{A}_{a_1\ldots a_m}:=(m-1)!\mathcal{F}_{(a_1\ldots a_m)}:=\frac{1}{m}\sum_{\sigma\in S_m}\eta(\sigma;a_1,\ldots a_m)\mathcal{F}_{a_{\sigma(1)}\ldots a_{\sigma(m)}}(t)
\end{equation}
Note that the parameters $s_i$ are super--commuting since those associated to even boundary states are anticommuting. The sign factor $\eta$ comes from permuting the boundary operators.
%%%%%%%%%%%%%%%%%%%%%%%%%%%%%%%%%%%%%%%%%%%%%%%%%%%%%%%%%%%%%%%%%%%%%%%%%%%%%%%%%%%%%
\subsection{Deformations and the Bulk--Boundary Crossing Constraint}
As noted before, we can not rely on Eq.~(\ref{eq:cardy}) for models other than the $A$-minimal models. Therefore some of the amplitudes will not be fixed by the $A_{\infty}$-- and the bulk--boundary crossing relations. 
We now give a prescription how to determine the values of all the correlators with bulk insertions by merging the generalized WDVV equations with the methods from section \ref{masseysec}. The procedure can be cast into the following recipe:
\begin{itemize}
\item Without bulk insertions, the bulk--boundary crossing constraint does not contain any extra information. This is why we may assume that in this case the superpotential coming from the $A_{\infty}$--relations and the one coming from the versal deformation of the $Q$--operator will agree. We thus start by computing the superpotentials for the bulk parameters set to zero with either method.
\item The superpotential obtained from solving the $A_{\infty}$--relations will contain as undetermined parameters non--linear functions of the unknown correlators. Comparing with the result of the Massey product algorithm one obtains an overdetermined system of non--linear equations for the correlators which has, at least for the examples we checked, a unique solution.
\item Next we set up the WDVV--constraints with the bulk parameters turned on and use the boundary correlators whose values we found by comparison of the two superpotentials as input for solving the equations. It turns out that this is enough to uniquely determine all the values of the remaining correlators and the complete superpotential is fixed up to reparameterizations. 
\end{itemize}

\subsection{Comparing Results}
For the $E_6$ case, we find
\begin{eqnarray}
\label{supobulkbbcr}
\overline{\mathcal{W}}_{CFT}(u;t)&=&u_4^3u_1+\frac{3}{4}u_4^2u_1^5+\frac{1}{8}u_4u_1^9+\frac{5}{832}u_1^{13} \nonumber\\
&&+\frac{1}{2}u_1\left(t_{12}-\frac{1}{2}t_6^2-\frac{1}{2}t_5^2t_2-\frac{1}{2}t_8t_2^2+\frac{1}{6}t_6t_2^3+\frac{1}{8}\left(t_6^2-t_6t_2^3+\frac{1}{4}t_2^6 \right) \right)\nonumber \\
&&+\frac{1}{2}  \left(u_4+\frac{1}{8}u_1^4\right)\left(t_9-t_5t_2^2\right)+\frac{1}{4}  \left(u_4u_1^3+\frac{3}{28}u_1^7\right)\left(t_6-\frac{1}{2}t_2^3\right).
\end{eqnarray}
From the point of view of the consistency constraints it is natural to use the ``flat'' parameters $t_i$ since the bulk--boundary crossing constraint contains the bulk prepotential $\mathcal{F}(t)$ \cite{Dijkgraaf:1990dj}. We can now make a field redefinition of the open string parameters in order to get maximum agreement with~(\ref{eq:fullpot}). This yields
\begin{equation}
\begin{array}{rcl}
\mathcal{W}_{CFT}(u;s)&=&\frac{5}{64\cdot 13}u_1^{13}+\frac{1}{8}u_4 u_1^9+\frac{3}{4}u_4^2u_1^5
+u_4^3 u_1+ u_1(s_{12}+\frac{1}{4}s_6^2)\\
&&+s_9 u_4+\frac{1}{8}s_9u_1^4+\frac{1}{2}s_6 u_4 u_1^3+\frac{3}{56}s_6 u_1^7,
\end{array}\label{eq:cftpot}
\end{equation}
where the $s_i$ are given in Eq.~(\ref{eq:si}).
Here only the $A_{\infty}$- and Crossing constraint were used, not the Cardy
equation. 
This solution corresponds to Eq.~(\ref{eq:fullpot}) with
some deformation parameters set to zero,
\begin{equation}
\mathcal{W}_{CFT}(u;s)=\mathcal{W}_{Massey}(u;s_2=0,s_5=0,s_6,s_8=0,s_9,s_{12}).\nonumber
\end{equation}
Thus, incorporating the bulk--boundary crossing constraint leads to a reduced version of the effective superpotential. This result suggests that the bulk--boundary crossing equations impose an additional constraint on the superpotential.\footnote{In technical terms, this means that the bulk--boundary crossing constraint sets certain correlators to zero which are left undetermined when only imposing the $A_{\infty}$ relations.} It might be that if one deforms the matrix factorization one only captures the $A_{\infty}$--structure and that one gets additional constraints from the interaction between bulk and boundary. This is incorporated in a mathematical structure termed Open Closed Homotopy Algebra (OCHA) which has recently been introduced in the literature \cite{Kajiura:2005sn}.\\
It is also  possible that missing terms in the bulk--boundary crossing constraint modify the relations such that one can get agreement with the result obtained from the Massey product method \cite{getzlerpriv}.  

%%%%%%%%%%%%%%%%%%%%%%%%%%%%%%%%%%%%%%%%%%%%%%%%%%%%%%%%%%%%%%%%%%%%%%%%%%%%%%%%%%%%%
\section{Further Results}
\label{moresec}
\subsection{The Exceptional Singularities}
In this section we list some more results for the superpotentials of the
minimal models of types $E_6,E_7,E_8$. For simplicity, all bulk parameters are set to 0.\\
First, we consider the the self--dual matrix factorization $M_3$ for $E_6$ given in
(\ref{m3}) where we choose $\varepsilon=-1$ in order to have only real entries in the
matrix. The fermionic spectrum can be read off from table (\ref{threevarcoh}).
There are four fermionic states and their deformation parameters $u_i$ have
charges $\{1,3,4,6\}$ and the four corresponding polynomials $f_i$ of the
ring $k[u_i]/(f_i)$ have charges $\{12,10,9,7\}$.
\begin{eqnarray}
f_1&=&-u_6^2+u_4^3-u_3^4-18u_4u_3^2u_1^2+7u_6u_3u_1^3-6u_4^2u_1^4-26u_3^2u_1^6-9u_4u_1^8-3u_1^{12}\nonumber \\
f_2&=&4u_3^3u_1+12u_4u_3u_1^3+8u_3u_1^7\nonumber \\
f_3&=&3u_4^2u_1-6u_3^2u_1^3-u_1^9\nonumber \\
f_4&=&-2u_6u_1+4u_3u_1^4
\end{eqnarray}
These polynomials can be integrated to the effective superpotential
\begin{equation}
\begin{array}{rcl}
\label{supom3}
\mathcal{W}_{eff}(u)&=&-u_6^2u_1+u_4^3u_1-u_3^4u_1-6u_4u_3^2u_1^3+4u_6u_3u_1^4\\
&&-8u_3^2u_1^7-u_4u_1^9-\frac{5}{13}u_1^{13}.
\end{array}
\end{equation}
Our last example for the $E_6$ model is the factorization $M_4$ given in Eq.~(\ref{m4}), where,
once again, we set $\varepsilon=-1$. We find the following expression for the
superpotential:
\begin{eqnarray}
\label{supom4}
\begin{array}{rcl}
\mathcal{W}_{eff}(u)&=&\frac{5}{1664}u_1^{13}-\frac{1}{4}u_1^8u_5+\frac{3}{4}u_1^6u_2u_5-2u_2^4u_5+2u_2u_3^2u_5-u_4^2u_5\\
&&-2u_3u_5^2+u_1^3\left(-u_2u_3u_5-\frac{1}{2}u_5^2\right)+u_1^9\left(-\frac{1}{16}u_2^2+\frac{1}{16}u_4+\frac{1}{16}v_4\right)\\
&&-u_4u_5v_4+u_5v_4^2+u_1^4\left(-\frac{3}{2}u_2^2u_5+u_4u_5+\frac{1}{2}u_5v_4\right)\\
&&+u_1^5\bigg(-\frac{3}{8}u_2^4-\frac{3}{8}u_4^2-\frac{1}{2}u_3u_5+u_2^2\left(\frac{3}{4}u_4+\frac{3}{4}v_4\right)-\frac{3}{4}u_4v_4-\frac{3}{8}v_4^2\bigg)\\
&&+u_1^2\left(2u_2^3u_5-u_3^2u_5+u_2\left(-2u_4u_5-u_5v_4\right)\right)+u_2^2\left(3u_4u_5+3v_4u_5\right)\\
&&+u_1\bigg(-\frac{1}{2}u_2^6+\frac{1}{2}u_4^3+u_2u_5^2+\frac{3}{2}u_4^2v_4+\frac{3}{2}u_4v_4^2+\frac{1}{2}v_4^3\\
&&+u_2^4\left(\frac{3}{2}u_4+\frac{3}{2}v_4\right)+u_3\left(u_4u_5+2u_5v_4\right)\\
&&+u_2^2\left(-\frac{3}{2}u_4^2+u_3u_5-3u_4v_4-\frac{3}{2}v_4^2\right)\bigg).\\
\end{array}
\end{eqnarray}
With (\ref{eq:fullpot}), (\ref{supom3}) and (\ref{supom4}) we have actually given the
superpotentials of five of the six branes since a brane yields the same superpotential
as its anti--brane.\\

For the $E_7$ singularity with $W=x^3+xy^3+z^2$ we consider its
simplest factorization, a self-dual one,
\begin{equation}
E=J=\left(\begin{array}{cc}
z&x\\
x^2+y^3&-z
\end{array}\right).
\end{equation}
The odd spectrum consists of three states with charges $\{0,8,16\}$ to which we associate parameters $u_i$ with charges $\{1,5,9\}$. The deformed matrix is given by:
\begin{eqnarray}
E_{def}&=&\bigg(\begin{array}{c}
z + u_1y^2 + u_5y + u_9\\
y^3 + x^2 - u_1^2x y - 2u_1u_5x + u_1^4y^2 + 4u_1^3u_5y - 8u_1^3u_9 + 
  u_1^6x - 2u_1^8y + 20u_1^7u_5 - 11u_1^{12}
\end{array}\nonumber\\
&&\begin{array}{c}
x + u_1^2y + 2u_1u_5 - u_1^6\\
-z + u_1y^2 + u_5y + u_9
\end{array}
\bigg)
\end{eqnarray}
The polynomials defining the deformation ring are:
\begin{eqnarray}
f_1&=&-u_9^2 - 16u_1^4u_5u_9 + 40u_1^8u_5^2 + 8u_1^9u_9 - 42u_1^{13}u_5 + 
    11u_1^{18}\nonumber \\
f_2&=&-2u_5u_9 + 8u_1^4u_5^2 - 8u_1^5u_9 + 12u_1^9u_5 - 9u_1^{14}\nonumber \\
f_3&=&-u_5^2 - 2u_1u_9 + 6u_1^5u_5 - 3u_1^{10}
\end{eqnarray}
These polynomials of degrees $\{18, 14, 10\}$ are easily integrated to the corresponding effective superpotential of degree 19,
\begin{equation}
\label{e7supo}
\mathcal{W}_{eff}(u)=-\frac{55}{19}u_1^{19}+12u_1^{14}u_5-15u_1^9u_5^2+5u_1^4u_5^3-3u_1^{10}u_9+6u_1^5u_5u_9-u_5^2u_9-u_1u_9^2.
\end{equation}
This can be related as follows to the coset model
\begin{equation}
\frac{E_7}{E_6\times U(1)}.\nonumber
\end{equation}
From~\cite{Eguchi:2001fm} we take the expression for the coset
Landau--Ginzburg potential:
\begin{equation}
\begin{array}{rcl}
W(x,y,z)&=&\frac{1016644}{817887699}x^{19}+\frac{33326}{177147}x^{14}y+\frac{266}{6561}x^{10}z+\frac{16850}{2187}x^9y^2\\
&&+\frac{80}{27}x^5yz+\frac{124}{9}x^4y^3+xz^2+\frac{3}{2}y^2z
\end{array}
\end{equation}
And indeed, this expression can be obtained from (\ref{e7supo}) by a field redefinition.\\

Finally we also give an example for the $E_8$ model. The Landau--Ginzburg potential
for this model is $W=x^3+y^5+z^2$ and its dual Coxeter number is $k=30$.
The simplest matrix factorization has rank $4$:
\begin{equation}
E=J=\left(\begin{array}{cccc}
z& 0& x& y\\
0& z& y^4& -x^2\\
x^2& y& -z& 0\\
y^4& -x& 0& -z
\end{array}
\right)
\end{equation}
The corresponding superpotential is
\begin{eqnarray}
\begin{array}{rcl}
\mathcal{W}_{eff}&=&-\frac{11}{31}u_1^{31}+u_1^{25}u_6-10u_1^{19}u_6^2+45u_1^{13}u_6^3-55u_1^7u_6^4-u_1u_6^5+3u_1^{21}u_{10}\\
&&-15u_1^{15}u_6u_{10}+15u_1^9u_6^2u_{10}+10u_1^3u_6^3u_{10}-u_1u_{10}^3-3u_1^{16}u_{15}\\
&&+10u_1^{10}u_6u_{15}-10u_1^4u_6^2u_{15}-u_1u_{15}^2+3u_{1}^{21}u_{10}-u_1u_{10}^3-3u_1^{16}u_{15}.
\end{array}
\end{eqnarray}
We cannot relate this result to a coset model, since such models do not exist
for $E_8$. This supports the conjecture that matrix factorizations of rank
greater than $2$ can not be related to such coset models.
%%%%%%%%%%%%%%%%%%%%%%%%%%%%%%%%%%%%%%%%%%%%%%%%%%%%%%%%%%%%%%%%%%%%%%%%%%%%%%%%%%%%
\section{Conclusions and Open Questions}
\label{conclusionsec}
In this paper we discussed various methods to calculate the effective superpotential for ADE minimal models with D--branes. The method of computing formal moduli provides a very efficient and elegant method to calculate the effective superpotential. It circumvents the problem that the Cardy condition as given in \cite{Herbst:2004jp} does not hold in the general case, which had been uncertain before. We were able to calculate various examples of effective superpotentials, most of which would have 
required an exceedingly high amount of computing time if tackled by implementing the
consistency constraints. Superpotentials of matrix factorizations whose spectrum contained more than 4 (even or odd) states were not calculated since computing time quickly increases as more states are added. The deformation calculus using the Massey products as implemented in {\sc Singular} allowed us to go further than that
and compute for example the potentials for all $E_6$ branes except the most complicated one, for which the calculation exceeded the powers of an ordinary PC.\\
As always, a number of open questions remain. One issue is the deformation of the matrix factorization by bosonic states. In the context of obstruction theory the fermionic states provide the deformations and the bosonic states give the obstructions. In this formalism the deformations with even states 
can not be computed. From the point of view of the CFT constraints, even and odd states are treated on equal footing, the only difference being that the even states are associated to anticommuting deformation parameters whereas the parameters for the odd states are commuting. We will leave a
possible generalization of the Massey product formalism to incorporate the even deformations as a future task.\\
An even more interesting question would be the generalization to Calabi-Yau manifolds which is of course the ultimate goal.
In principle the computation should proceed just as in the ADE-case but then we also have marginal deformations. The results of \cite{Ashok:2004xq} may be useful for a better understanding of this problem.\\
In this paper we dealt only with the boundary preserving sector but
it would be interesting to consider systems with various D--branes.
Some rather na\"ive experimentation with the A--minimal models leads to the conclusion that a
generalization of the algorithm to the case of multiple D--branes
is straightforward although the technical complexity grows quickly. It seems likely
that this is a powerful framework to describe phenomena like tachyon
condensation and bound state formation of D--branes.\\
Another interesting issue is the relation to coset models. We have been very brief
about the relations between the superpotentials and we only gave an idea of how the
matrix factorizations could be recovered from these models. It may be useful to
investigate this relation further. In particular, one should try to find out
whether also matrix factorizations of rank higher than two have connections to
coset models.\\
Last but not least it may be interesting to look for terms missing in the Cardy condition and in the bulk--boundary crossing constraint. \\\\
%%%%%%%%%%%%%%%%%%%%%%%%%%%%%%%%%%%%%%%%%%%%%%%%%%%%%%%%%%%%%%%%%%%%%%%%%%%%%%%
\textbf{Acknowledgements} We are deeply grateful to our advisor Wolfgang Lerche
for introducing us to the topic of this paper and for many useful explanations and enlightening discussions.\\
We would also like to thank Gerhard Pfister, Johannes Walcher and Nicholas Warner
for helpful comments. \\
Furthermore we thank Ezra Getzler and Andreas Recknagel for valuable discussions and useful comments.
%%%%%%%%%%%%%%%%%%%%%%%%%%%%%%%%%%%%%%%%%%%%%%%%%%%%%%%%%%%%%%%%%%%%%%%%%%%%%%%%%%%%
\appendix
\section{$E_6$ -- Three--Variable Case}
\subsection{Matrix Factorizations}
\label{threevar-app}
These results were already given in \cite{Kajiura:2005yu}. The superpotential is:\begin{equation}
W=x^3+y^4+\varepsilon\,z^2,
\end{equation} 
where $\varepsilon=\pm 1$. Introducing this parameter is just for calculational convenience since we can always choose the matrix factorization to be real. The choice of sign has no influence on the dimensions and charges of the spectrum or the form of the superpotential. The following matrices satisfy $W=E_i\cdot J_i$:
\begin{equation}
E_1=J_2=\left(
\begin{array}{cc}
-y^2+\sqrt{-\varepsilon}z&x\\
x&y^2+\sqrt{-\varepsilon}z
\end{array}
\right)\quad
E_2=J_1=\left(
\begin{array}{cc}
-y^2-\sqrt{-\varepsilon}z&x\\
x&y^2-\sqrt{-\varepsilon}z
\end{array}
\right)
\end{equation}
\begin{equation}
\label{m3}
E_3=J_3=\left(
\begin{array}{cccc}
-\sqrt{\varepsilon}z&0&x^2&y^3\\
0&-\sqrt{\varepsilon}z&y&-x\\
x&y^3&\sqrt{\varepsilon}z&0\\
y&-x^2&0&-\sqrt{\varepsilon}z
\end{array}
\right)
\end{equation}
\begin{equation}
\label{m4}
E_4=J_5=\left(
\begin{array}{cccc}
-y^2+\sqrt{-\varepsilon}&0&xy&x\\
-xy&y^2+\sqrt{-\varepsilon}&x^2&0\\
0&x&\sqrt{-\varepsilon}z&y\\
x^2&-xy&y^3&\sqrt{-\varepsilon}z
\end{array}
\right)
\end{equation}
\begin{equation}
E_5=J_4=\left(
\begin{array}{cccc}
-y^2-\sqrt{-\varepsilon}z&0&xy&x\\
-xy&y^2-\sqrt{-\varepsilon}z&x^2&0\\
0&x&-\sqrt{-\varepsilon}z&y\\
x^2&-xy&y^3&-\sqrt{-\varepsilon}z
\end{array}
\right)
\end{equation}
\begin{equation}
E_6=\left(
\begin{array}{cccccc}
-\sqrt{-\varepsilon}z&-y^2&xy&0&x^2&0\\
-y^2&-\sqrt{-\varepsilon}z&0&0&0&x\\
0&0&-\sqrt{-\varepsilon}z&-x&0&y\\
0&xy&-x^2&-\sqrt{-\varepsilon}z&y^3&0\\
x&0&0&y&-\sqrt{-\varepsilon}z&0\\
0&x^2&y^3&0&xy^2&-\sqrt{-\varepsilon}z
\end{array}
\right)
\end{equation}
\begin{equation}
J_6=\left(
\begin{array}{cccccc}
\sqrt{-\varepsilon}z&-y^2&xy&0&x^2&0\\
-y^2&\sqrt{-\varepsilon}z&0&0&0&x\\
0&0&\sqrt{-\varepsilon}z&-x&0&y\\
0&xy&-x^2&\sqrt{-\varepsilon}z&y^3&0\\
x&0&0&y&\sqrt{-\varepsilon}z&0\\
0&x^2&y^3&0&xy^2&\sqrt{-\varepsilon}z
\end{array}
\right)
\end{equation}

\subsection{Spectrum}
\label{e6threesec}
The spectrum for this model has already been discussed in \cite{Kajiura:2005yu}. There are six matrix factorizations, one for each node in the Dynkin diagram. We summarize the data of the boundary preserving spectrum in the following table:
\begin{equation}
\label{threevarcoh}
\begin{tabular}{|c|c|c|c|}
\hline
\vrule width 0pt height 12pt depth 6ptFactorization& Rank & Spectrum bosonic & Spectrum fermionic\\ \hline
\vrule width 0pt height 12pt depth 6pt $M_1$ & $2$ & $0\:6$ & $4\:10$\\ \hline
\vrule width 0pt height 12pt depth 6pt $M_2$ & $2$ & $0\:6$ & $4\:10$\\ \hline
\vrule width 0pt height 12pt depth 6pt $M_3$ & $4$ & $0\:4\:6\:10$ & $0\:4\:6\:10$\\ \hline
\vrule width 0pt height 12pt depth 6pt $M_4$ & $4$ & $0\:2\:4\:6^2\:8$ & $2\:4^2\:6\:8\:10$\\ \hline
\vrule width 0pt height 12pt depth 6pt $M_5$ & $4$ & $0\:2\:4\:6^2\:8$ & $2\:4^2\:6\:8\:10$\\ \hline 
\vrule width 0pt height 12pt depth 6pt $M_6$ & $6$ & $0\:2^2\:4^3\:6^3\:8^2\:10$ & $0\:2^2\:4^3\:6^3\:8^2\:10$ \\
\hline
\end{tabular}
\end{equation}
We labelled the matrix factorizations by $M_i$, the second column indicates the ranks of the matrices. The last two columns give the even and the odd spectrum. The numbers correspond to the $R$--charges multiplied by the number $12$ -- the Coxeter number of $E_6$ -- and the exponents give the multiplicities. Note that there are six possible values of the charges, $q_{\psi}\in\{0,2,4,6,8,10\}$. To fermionic states with these charges we associate fermionic deformation parameters $u_i$ with charges $q_{u_i}=\frac{1}{2}(12-q_{\psi_i})\in\{6,5,4,3,2,1\}$.  We observe that, concerning the spectrum, there are two types of matrix factorizations. The factorizations $M_1,M_2$ and $M_4,M_5$ have the same spectra, respectively. These branes are the antibranes of each other. $M_3$ and $M_6$ belong to a different class of D--branes. The even spectrum is identical to the odd spectrum, these branes are ``self--dual'' ---  the brane is its own antibrane \cite{Kapustin:2003rc}. \\
We observe that the highest charge, which is equal to the background charge, is always in the fermionic sector, whereas the charge $0$ state is always in the bosonic sector. In fact, it is possible to determine the degree of the effective superpotential just by charge considerations. Remember the selection rules for the correlators given in section \ref{selection-subsec}. The allowed correlators must have an $R$--charge which is equal to the background charge $q_b$. We can now determine the correlator with the maximal number of insertions. This correlator will have three unintegrated insertions of the field with the highest charge $q_b$, we will call this field $\psi_b$, and a certain number of integrated operators, which have negative charge. To get the maximum number of insertions one must use only insertions of $\int\psi_b$, which has charge $b-1$, which is the least negative. From the charge selection rule we can now calculate the the number $x$ of integrated insertions of the top element:
\begin{equation}
3\cdot b+x\cdot(b-1)\stackrel{!}{=}b
\end{equation}
This yields $x=\frac{2b}{1-b}$. Now take into account that for the minimal models the background charge is related to the Coxeter number $k$ via $b=\frac{k-2}{k}$. Inserting this, we find that the number of integrated insertions of the top--element is $k-2$. Adding the three unintegrated operators, one finds that the top--correlator is a $k+1$--point function. Looking more closely, one also finds that the selection rule for the $\mathbb{Z}_2$--charge is satisfied and that this correlator will not vanish and contribute to the superpotential. The deformation parameter $u$ associated to $\psi_b$ has charge one and we will get a term $u_1^{k+1}$ in the superpotential. Since the effective superpotential is a homogeneous polynomial, we conclude:\\

{\it The effective superpotentials for the ADE minimal models always have degree $k+1$, where $k$ is the Coxeter number.}

\section{$E_6$ -- Two--Variable Case}
There is also a two--variable description for the $E_6$ minimal model:
\begin{equation}
W=x^3+y^4
\end{equation}
For completeness we list all the matrix factorizations and give
the complete spectrum. From the point of view of conformal field theory these
two incarnations of the $E_6$--model correspond to two different GSO projections
\cite{Kapustin:2003rc,Brunner:2005pq,Brunner:2005fv}.

\subsection{Matrix Factorizations}
We find the following matrix factorizations:
\begin{equation}
E_1=J_2=\left(\begin{array}{cc}x&y\\y^3&-x^2\end{array}\right)\quad
E_2=J_1=\left(\begin{array}{cc}x^2&y\\y^3&-x\end{array}\right)
\end{equation}
\begin{equation}
E_3=\left(\begin{array}{cc}x&y^2\\y^2&-x^2\end{array}\right)\qquad
J_3=\left(\begin{array}{cc}x^2&y^2\\y^2&-x\end{array}\right)
\end{equation}
\begin{equation}
E_4=J_5=\left(\begin{array}{ccc}
x&y&0\\
0&x&y\\
y^2&0&x
\end{array}\right)
\quad
E_5=J_4=\left(\begin{array}{ccc}
x^2&-xy&y^2\\
y^3&x^2&-xy\\
-xy^2&y^3&x^2
\end{array}\right)
\end{equation}
\begin{equation}
E_6=\left(
\begin{array}{cccc}
x&y^2&0&0\\
y^2&-x^2&0&0\\
0&-xy&x^2&y^2\\
y&0&y^2&-x
\end{array}
\right)
\quad
J_6=\left(
\begin{array}{cccc}
x^2&y^2&0&0\\
y^2&-x&0&0\\
0&-y&x&y^2\\
xy&0&y^2&-x^2
\end{array}
\right)
\end{equation}

\subsection{Spectrum}
For the even spectrum we find:
\begin{equation}
\label{twovar-bosonic}
\begin{tabular}{|c|c|c|c|c|c|c|}\hline
\vrule width 0pt height 12pt depth 6pt & $1$&$2$&$3$&$4$&$5$&$6$\\ \hline
\vrule width 0pt height 12pt depth 6pt $1$&$0\:10$&$4\:6$&$3\:7$&$1\:3\:5$&$5\:7\:9$&$2\:4\:6\:8$\\ \hline
\vrule width 0pt height 12pt depth 6pt $2$&$4\:6$&$0\:10$&$3\:7$&$5\:7\:9$&$1\:3\:5$&$2\:4\:6\:8$\\ \hline
\vrule width 0pt height 12pt depth 6pt $3$&$3\:7$&$3\:7$&$0\:4\:6\:10$&$2\:4\:6\:8$&$2\:4\:6\:8$&$1\:3\:5^2\:7\:9$\\ \hline
\vrule width 0pt height 12pt depth 6pt $4$&$5\:7\:9$&$1\:3\:5$&$2\:4\:6\:8$&$0\:2\:4\:6\:8\:10$&$2\:4^2\:6^2\:8$&$1\:3^2\:5^2\:7^2\:9$\\ \hline
\vrule width 0pt height 12pt depth 6pt $5$&$1\:3\:5$&$5\:7\:9$&$2\:4\:6\:8$&$2\:4^2\:6^2\:8$&$0\:2\:4\:6\:8\:10$&$1\:3^2\:5^2\:7^2\:9$\\ \hline
 \vrule width 0pt height 12pt depth 6pt $6$&$2\:4\:6\:8$&$2\:4\:6\:8$&$1\:3\:5^2\:7\:9$&$1\:3^2\:5^2\:7^2\:9$&$1\:3^2\:5^2\:7^2\:9$&$0\:2^2\:4^3\:6^3\:8^2\:10$ \\ \hline
\end{tabular}
\end{equation}
The odd spectrum is summarized in the following table:
\begin{equation}
\begin{tabular}{|c|c|c|c|c|c|c|}\hline
\vrule width 0pt height 12pt depth 6pt & $1$&$2$&$3$&$4$&$5$&$6$\\ \hline
\vrule width 0pt height 12pt depth 6pt $1$&$4\:6$&$0\:10$&$3\:7$&$5\:7\:9$&$1\:3\:5$&$2\:4\:6\:8$\\ \hline
\vrule width 0pt height 12pt depth 6pt $2$&$0\:10$&$4\:6$&$3\:7$&$1\:3\:5$&$5\:7\:9$&$2\:5\:6\:8$\\ \hline
\vrule width 0pt height 12pt depth 6pt $3$&$3\:7$&$3\:7$&$0\:4\:6\:10$&$2\:4\:6\:8\:$&$2\:4\:6\:8$&$1\:3\:5^2\:7\:9$\\ \hline
\vrule width 0pt height 12pt depth 6pt $4$&$1\:3\:5$&$5\:7\:9$&$2\:4\:6\:8$&$2\:4^2\:6^2\:8$&$0\:2\:4\:6\:8\:10$&$1\:3^5\:5^2\:7^2\:9$\\ \hline
\vrule width 0pt height 12pt depth 6pt $5$&$5\:7\:9$&$1\:3\:5$&$2\:4\:6\:8$&$0\:2\:4\:6\:8\:10$&$2\:4^2\:6^2\:8$&$1\:3^2\:5^2\:7^2\:9$\\ \hline
 \vrule width 0pt height 12pt depth 6pt $6$&$2\:4\:6\:8$&$2\:4\:6\:8$&$1\:3\:5^2\:7\:9$&$1\:3^2\:5^2\:7^2\:9$&$1\:3^2\:5^2\:7^2\:9$&$0\:2^2\:4^3\:6^3\:8^2\:10$ \\ \hline
\end{tabular}
\end{equation}

\section{Some Results for the $A$--Series}
\label{cardy-app}
The minimal models of the $A$--Series satisfy an additional constraint, the Cardy Condition \cite{Herbst:2004jp}:
\begin{equation}
\partial_i\mathcal{F}_{a_0\ldots a_n}\eta^{ij}\partial_j\mathcal{F}_{b_0\ldots b_m}=
\end{equation}
\[
\sum_{\stackrel{{0\leq n_1\leq n_2\leq n}}{0\leq m_1\leq m_2\leq m}}(-1)^{(\tilde{c}_1+\tilde{a}_0)(\tilde{c}_2+\tilde{b}_0)+\tilde{c_1}+\tilde{c_2}}\omega^{c_1d_1}\omega^{c_2d_2}\mathcal{F}_{a_0\ldots a_{n_1}d_1b_{m_1+1}\ldots m_2c_2a_{n_2+1}\ldots a_n}\mathcal{F}_{b_0\ldots b_{m_1}c_1a_{n_1+1}\ldots a_{n_2}d_2b_{m_2+1}\ldots b_m}
\]
Setting $t=0$ and $m=0,n=0$ one recovers the CFT Cardy constraint, which is satisfied for all the minimal models. The full constraint in the form given above is only satisfied for the $A$--Series. For the generic case it has to be replaced by the Quantum $A_{\infty}$ Structure \cite{Herbst:2006kt,Herbst:2006nn}.
The Cardy constraint fixes the reparameterization freedom of the superpotential. The superpotential for the $A_k$--model is:
\begin{equation}
W^{(k+2)}(x)=\frac{x^{k+2}}{k+2},
\end{equation}
where the exponents in brackets give the degrees of the polynomials.
The matrix factorizations are:
\begin{equation}
W^{k+2}(x)=E^{\kappa+1}(x)J^{k+1-\kappa}(x),\qquad\qquad\kappa=0,\ldots,[k/2]
\end{equation}
We denote by $h$ the greatest common denominator of $E$ and $J$, i.e $E^{\kappa+1}(x)=p(x)h^{\ell+1}(x)$ and $J^{k+1-\kappa}(x)=q(x)h^{\ell+1}(x)$. The pair $(k,\ell)$ then uniquely labels the D--brane we consider. 
For the $A_3$--model with $W=\frac{x^5}{5}$ and $(k,\ell)=(3,1)$ only one incarnation of the superpotential satisfies the Cardy constraint because this condition fixes the reparameterization freedom of the effective superpotential \cite{Herbst:2004jp}:
\begin{eqnarray}
\mathcal{W}_{eff}(t,u)&=&-\frac{1}{5}\left(\frac{u_1^6}{6}+u_1^4u_2+\frac{3}{2}u_1^2u_2^2+\frac{u_2^3}{3}\right)-t_2\left(-\frac{u_1^4}{4}-u_1^2u_2-\frac{u_2^2}{2}\right)+t_3\left(\frac{u_1^3}{3}+u_2u_2\right)\nonumber\\
&&-\left(t_4-t_2^2\right)\left(-\frac{u_1^2}{2}-u_2\right)+\left(t_5-t_2t_3\right)u_1
\end{eqnarray}
It turns out that the Cardy constraint is only valid in the boundary preserving sector when bulk perturbations are turned on. But this information is enough to obtain superpotentials for the boundary changing sector. We now state some new results for superpotentials in the boundary changing sector. For the model $(3,1)\oplus (3,0)$, we find:
\begin{eqnarray}
\mathcal{W}_{eff}&=&-\frac{1}{5}\bigg(\frac{u_1^6}{6}+\frac{w_1^6}{6}+u_1^4u_2+\frac{3}{2}u_2^2u_1^2+\frac{u_2^3}{3}+2u_2u_1v_{\frac{3}{2}}\tilde{v}_{\frac{3}{2}}+u_1^3v_{\frac{3}{2}}\tilde{v}_{\frac{3}{2}}+\frac{1}{2}v_{\frac{3}{2}}^2\tilde{v}_{\frac{3}{2}}^2+u_2v_{\frac{3}{2}}\tilde{v}_{\frac{3}{2}}w_1\nonumber\\
&&+u_1^2v_{\frac{3}{2}}\tilde{v}_{\frac{3}{2}}w_1+u_1v_{\frac{3}{2}}\tilde{v}_{\frac{3}{2}}w_1^2+v_{\frac{3}{2}}\tilde{v}_{\frac{3}{2}}w_1^3\bigg)-t_2\left(-\frac{u_1^4}{4}-\frac{w_1^4}{4}-u_1^2u_2-\frac{u_2^2}{2}-u_1v_{\frac{3}{2}}\tilde{v}_{\frac{3}{2}}-v_{\frac{3}{2}}\tilde{v}_{\frac{3}{2}}w_1\right)\nonumber\\
&&+t_3\left(\frac{u_1^3}{3}+\frac{w_1^3}{3}+u_1u_2+v_{\frac{3}{2}}\tilde{v}_{\frac{3}{2}}\right)-\left(t_4-t_2^2\right)\left(-\frac{u_1^2}{2}-\frac{w_1^2}{2}-u_2\right)+\left(t_5-t_2t_3\right)\left(-u_1-w_1\right)\nonumber\\
\end{eqnarray}
Here we chose the convention that the indices correspond to the charges, the parameters $u$ and $w$ are related to odd states on the branes labeled with $\ell=1$ and $\ell=0$, respectively, whereas the parameters $v$ and $\tilde{v}$ are related to the open string states stretching between the two branes.\\
For the model $(4,1)\oplus(4,0)$ we find:
\begin{eqnarray}
\mathcal{W}_{eff}&=&-\frac{1}{6}\bigg(\frac{u_1^7}{7}+\frac{w_1^7}{7}+2u_1^3u_2^2+u_1^5u_2+u_1u_2^3+u_1^4v_{\frac{3}{2}}\tilde{v}_{\frac{3}{2}}+u_1^3v_{\frac{3}{2}}\tilde{v}_{\frac{3}{2}}w_1+3u_1^2u_2v_{\frac{3}{2}}\tilde{v}_{\frac{3}{2}}+u_1^2v_{\frac{3}{2}}\tilde{v}_{\frac{3}{2}}w_1\nonumber\\
&&+u_1v_{\frac{3}{2}}^2\tilde{v}_{\frac{3}{2}}^2+u_1v_{\frac{3}{2}}\tilde{v}_{\frac{3}{2}}w_1^3+u_2^2v_{\frac{3}{2}}\tilde{v}_{\frac{3}{2}}+2u_2u_1v_{\frac{3}{2}}\tilde{v}_{\frac{3}{2}}w_1+u_2v_{\frac{3}{2}}\tilde{v}_{\frac{3}{2}}w_1^2+v_{\frac{3}{2}}^2\tilde{v}_{\frac{3}{2}}w_1+v_{\frac{3}{2}}\tilde{v}_{\frac{3}{2}}w_1^4\bigg)\nonumber\\
&&+t_2\left(\frac{u_1^5}{5}+\frac{w_1^5}{5}+u_1^3u_2+u_1^2v_{\frac{3}{2}}\tilde{v}_{\frac{3}{2}}+u_1u_2^2+u_1v_{\frac{3}{2}}\tilde{v}_{\frac{3}{2}}w_1+u_2v_{\frac{3}{2}}\tilde{v}_{\frac{3}{2}}+v_{\frac{3}{2}}\tilde{v}_{\frac{3}{2}}w_1^2\right)\nonumber\\
&&+t_3\left(\frac{u_1^4}{4}+\frac{w_1^4}{4}+u_1^2u_2+\frac{u_2^2}{2}+u_1v_{\frac{3}{2}}\tilde{v}_{\frac{3}{2}}+v_{\frac{3}{2}}\tilde{v}_{\frac{3}{2}}w_1\right)+\left(t_4-\frac{3}{2}t_2^2\right)\left(\frac{u_1^3}{3}+\frac{w_1^3}{3}+u_1u_2+v_{\frac{3}{2}}\tilde{v}_{\frac{3}{2}}\right)\nonumber\\
&&+\left(t_5-2t_2t_3\right)\left(\frac{u_1^2}{2}+\frac{w_1^2}{2}+u_2\right)+\left(t_6-\frac{1}{2}t_3^2-t_2t_4+\frac{1}{3}t_2^3\right)\left(u_1+w_1\right)
\end{eqnarray}
Finally, we give the result for $(4,0)\oplus(4,0)$
\begin{eqnarray}
\mathcal{W}_{eff}&=&-\frac{1}{6}\bigg(\frac{u_1^7}{7}+\frac{w_1^7}{7}+u_1^5v_1\tilde{v}_8+u_1^4v_1\tilde{v}_8w_1+2u_1^3v_1^2\tilde{v}_8^2+u_1^3v_1\tilde{v}_8w_1^2+u_1^2v_1\tilde{v}_8w_1^3+3u_1^2v_1^2\tilde{v}_8^2w_1\nonumber\\
&&+u_1v_1^3\tilde{v}_8^3+3u_1v_1^2\tilde{v}_8w_1^2+u_1v_1\tilde{v}_8w_1^4+v_1\tilde{v}_8w_1^5+2v_1^2\tilde{v}_8^2w_1^3+v_1^3\tilde{v}_8^3w_1\bigg)\nonumber\\
&&+t_2\left(\frac{u_1^5}{5}+\frac{w_1^5}{5}+u_1^3v_1\tilde{v}_8+u_1^2v_1\tilde{v}_8w_1+u_1v_1^2\tilde{v}_8^2+u_1v_1\tilde{v}_8w_1^2+v_1^2\tilde{v}_8^2w_1+v_1\tilde{v}_8w_1^3\right)\nonumber\\
&&+t_3\left(\frac{u_1^4}{4}+\frac{w_1^4}{4}+u_1^2v_1\tilde{v}_8+u_1v_1\tilde{v}_8w_1+\frac{1}{2}v_1^2\tilde{v}_8^2+v_1\tilde{v}_8w_1^2\right)\nonumber\\
&&+\left(t_4-\frac{3}{2}t_2^2\right)\left(\frac{u_1^3}{3}+\frac{w_1^3}{3}+u_1v_1\tilde{v}_8+v_1\tilde{v}_8w_1\right)+\left(t_5-2t_2t_3\right)\left(\frac{u_1^2}{2}+\frac{w_1^2}{2}+v_1\tilde{v}_8\right)\nonumber\\
&&+\left(t_6-\frac{1}{2}t_3^2-t_2t_4+\frac{1}{3}t_2^3\right)\left(u_1+w_1\right)
\end{eqnarray}
All of these results can be obtained from a residue formula \cite{Herbst:2004zm}:
\begin{equation}
\mathcal{W}_{eff}(t;u)=-\oint\frac{\mathrm{d}x}{2\pi i}\log(\det J(x;u))W(x;t),
\end{equation}
where $J(x;u)$ is the matrix factorization of the system with linear odd deformations turned on and $W(x;t)$ is the bulk superpotential \cite{Dijkgraaf:1990dj}. Note that this equation cannot be generalized to the multi--variable case in a straight forward manner. The fact that only linear deformations in the boundary parameters appear in $J(x;u)$ is a peculiarity of the $A_k$--models and does not hold in general. A possible hint for the generalization of this formula may be found in \cite{Eguchi:2001fm} where it was shown that in the context of coset models it is possible to obtain the superpotentials as period integrals over Calabi--Yau 4--folds which are fibrations of ALE singularities. In may prove worthwhile to investigate these results from the point of view of D--branes and matrix factorizations. 

%\bibliographystyle{utphys}
%\bibliography{bibliography}

\begingroup\raggedright\endgroup

\end{document}